\documentclass[a4paper,11pt]{article}
\usepackage{jcappub}
\usepackage[T1]{fontenc}
\usepackage{graphicx}
\usepackage{orcidlink}
\usepackage{amsmath}
\usepackage{xspace}
\usepackage[tablesonly,nomarkers,nolists,noheads]{endfloat}
\usepackage{physics}

\newcommand{\lat}{\textit{Fermi}-LAT\xspace}
\newcommand{\hess}{H.E.S.S.\xspace}

\begin{document}

\title{\boldmath Constraining cosmological parameters from very-high-energy $\gamma$-ray attenuation in blazar spectra: Bayesian inference and systematic uncertainties}

\author[a,b]{Patryk Liniewicz$^{\orcidlink{0009-0008-3575-3965}}$}
\author[c]{Emmanuel Moulin$^{\orcidlink{0000-0003-4007-0145}}$}

\affiliation[a]{Obserwatorium Astronomiczne, Uniwersytet Jagielloński, ul. Orla 171, 30-244 Kraków, Poland}
\affiliation[b]{Uniwersytet Jagielloński, Szkoła Doktorska Nauk Ścisłych i Przyrodniczych, ul. Prof. St. Łojasiewicza 11, 30-348 Kraków, Poland}

\affiliation[c]{Irfu, CEA Saclay, Universit\'e Paris-Saclay, F-91191 Gif-sur-Yvette, France}

\emailAdd{patryk.liniewicz@doctoral.uj.edu.pl}
\emailAdd{emmanuel.moulin@cea.fr}

\abstract{Very-high-energy $\gamma$ rays from active galactic nuclei are attenuated through pair production on the extragalactic background light (EBL), providing a probe of cosmic expansion. We develop a Bayesian inference analysis framework that uses an observationally driven, redshift-dependent EBL model and anchors intrinsic source spectra to the Fermi-LAT 3FHL catalogue while marginalising spectral-shape departures and dataset normalisations. The method is validated on realistic mock observations and applied to 241 archival H.E.S.S., MAGIC and VERITAS spectra of 50 AGN compiled in STeVECat. Because the leading opacity depends on the ratio of the EBL normalisation to the reduced Hubble constant, the gamma-ray data primarily constrain this ratio rather than $H_0$ alone. We measure $r=1.723^{+0.096}_{-0.095}$. For a fixed EBL scale, this gives $H_0=58.3^{+3.4}_{-3.0},\mathrm{km,s^{-1},Mpc^{-1}}$; allowing an 18\% EBL-scale uncertainty gives $H_0=60.5^{+10.8}_{-10.0},\mathrm{km,s^{-1},Mpc^{-1}}$. Matched injections recover the input opacity without appreciable bias, and the tested intrinsic-spectrum and EBL-shape perturbations are subdominant except for large infrared-background mismodelling. The predominantly low-redshift sample leaves $\Omega\mathrm{M}$ weakly constrained. The measurement is therefore limited by the EBL normalisation rather than statistics; reducing its uncertainty to a few per cent would make gamma-ray opacity a competitive, distance-ladder-independent probe of $H_0$.}


\maketitle

\section{Introduction}
Very-high-energy (VHE, $E \gtrsim 100$\,GeV) $\gamma$ rays travelling over cosmological distances interact with the Extragalactic Background Light (EBL) through pair production~\cite{Nikishov:1962rmq,Gould:1966pza}, which yields an attenuation in the VHE spectra of $\gamma$-ray sources. This effect can be characterised by an optical depth and has been measured by current atmospheric Cherenkov telescopes. The EBL is composed of redshifted photons produced by star formation processes over cosmic history~\cite{Hauser:2001xs}. This diffuse radiation field is accumulated in the intergalactic space from ultraviolet to infrared wavelengths.
The amount of attenuation along the line of sight depends on the EBL spectral energy density and the local expansion rate $H_0$ together with the cosmological parameters such as the matter content of the Universe $\Omega_\mathrm{M}$. A determination of $H_0$ and $\Omega_\mathrm{M}$ can be obtained by comparing the observed VHE $\gamma$-ray spectra from distant sources such as AGN to the expected ones with the attenuation from a given EBL model. 

An accurate determination of the EBL spectral energy density is required to recover the intrinsic properties of blazars~\cite{Dominguez:2015ama,Rosillo:2022xvr}, to understand the propagation of VHE $\gamma$ rays over cosmological distances~\cite{Aharonian:1993vz,DeAngelis:2007dqd,Broderick:2011av,Buehler:2020qsn}, to derive local cosmological parameters~\cite{Barrau:2008pi,Dominguez:2013mfa,Biteau:2015xpa,Dominguez:2019jqc}, to predict the detectability of sources with VHE $\gamma$-ray observatories from GeV observations, e.g., from \lat\ measurements~\cite{Paiano:2021jsa}, and to search for axion-like particles~\cite{Porras-Bedmar:2024uql,Grimaud:2026hpu}.
Comparing the $\gamma$-ray attenuation expected from EBL models with observations enables a determination of the cosmological distance as a function of redshift, and therefore a measurement of the Hubble constant. As we will see later, the pair-production cross section peaks at infrared wavelengths for TeV $\gamma$ rays; therefore, an accurate determination of the cosmic infrared background is required to reach the highest sensitivity in the derivation of the cosmological parameters.

In light of the persistent discrepancy
between the Hubble constant inferred from Cosmic Microwave Background (CMB) observations and the more local measurements from Cepheids and type Ia supernovae~\cite{Riess:2021jrx,Kamionkowski:2022pkx}, the so-called Hubble tension, independent determinations of $H_0$ that rely on different physics and carry different systematic uncertainties are particularly valuable. The $\gamma$-ray opacity of the Universe offers such a route: the EBL attenuation imprinted on VHE blazar spectra depends on the cosmological distance scale, and previous analyses have exploited it to derive $H_0$~\cite{Barrau:2008pi,Dominguez:2013mfa,Biteau:2015xpa,Dominguez:2019jqc,2024MNRAS.527.4632D}. In this work we revisit the method with a deliberately conservative treatment of the dominant systematic uncertainties. On the experimental side, source-level intrinsic indices and curvatures are hierarchically anchored to GeV measurements rather than freely fitted, blazar variability is absorbed by marginalised per-dataset flux normalisations.  On the modelling side, the EBL-scale uncertainty is being kept explicit by inferring the identifiable opacity ratio $r \equiv \alpha_\mathrm{EBL}/h$, with the adequacy of this modelling validated by dedicated mismodelling studies on mock data, and the impact of spectral uncertainties of the EBL SED is quantified. 
We eventually assess the performance of the determination of $r$ and the matter density $\Omega_\mathrm{M}$ in light of the above-mentioned uncertainties, and translate $r$ into $H_0$ only under explicitly stated external assumptions on the EBL scale.

This work is complementary to the most recent determinations using $\gamma$-ray opacity. The analysis output of the present work is twofold.
Firstly, it makes use of mock datasets in order to assess the sensitivity reach to cosmological parameters in light of a range of systematic uncertainties from both the modelling and experimental sides. Second, using STeVECat catalogued spectra of extragalactic VHE sources, we provide a determination of cosmological parameters according to conservative and optimistic sets of considered systematic uncertainties.
Gréaux et al.~\cite{Greaux:2024coc}, using STeVECat spectra, reconstruct the EBL spectral intensity itself modelled as eight Gaussians in log-wavelength with one evolution parameter, simultaneously with the intrinsic source spectra anchored to \lat\ priors, and report $H_0 \propto (1+f_\mathrm{diff})$, where $f_\mathrm{diff}$ is a possible diffuse component on top of the integrated galaxy light. Domínguez et al.~\cite{Dominguez:2019jqc,2024MNRAS.527.4632D} instead infer cosmology from previously published optical-depth measurements combined with galaxy-photometry EBL models. The present analysis takes a different approach using a Bayesian inference framework: the EBL spectral shape is extracted from the state-of-the-art, $\gamma$-ray-independent galaxy-photometry model of Salda\~na-L\'opez et al.~\cite{2021MNRAS.507.5144S} (hereafter SL21), its scale is kept explicit through the identifiable ratio $r=\alpha_\mathrm{EBL}/h$,
deviations in its SED are explored according to present uncertainties in its determination, and source-level index and curvature deviations from the GeV-band catalogue anchors are marginalised hierarchically.

The paper is organised as follows. Section~\ref{sec:ebl} summarises the latest measurements of the EBL spectral density, outlining the main features of an up-to-date EBL model built on a large sample of multiwavelength observations of galaxies, together with the derivation of the optical depth to $\gamma$ rays from a given redshift-dependent EBL model. In Section~\ref{sec:selection}, we perform a selection of AGN from GeV $\gamma$-ray measurements relevant for VHE $\gamma$-ray observations. Section~\ref{sec:analysis} is devoted to the statistical analyses of archival and mock combined observations of AGN, respectively, with imaging atmospheric Cherenkov telescopes (IACTs) to derive the Hubble constant and matter density in the local Universe, formulated directly in terms of the identifiable opacity ratio $r=\alpha_\mathrm{EBL}/h$. In Section~\ref{sec:results} the results are presented together with a discussion on the impact of systematic uncertainties from the modelling and the data. Section~\ref{sec:summary} is devoted to the conclusions of this work.

\section{Extragalactic background light and \texorpdfstring{$\gamma$}{gamma}-ray horizon}
\label{sec:ebl}
\subsection{Measurements and determination}
\label{subsec:meas}

The Extragalactic Background Light is a faint, diffuse radiation field spanning from infrared (IR) to ultraviolet (UV) wavelengths, produced predominantly by star formation processes integrated over the cosmic history. It thus contains imprints of both the expansion history of the Universe and the reprocessing of starlight via absorption and re-emission on dust in galaxies. Detailed knowledge of galaxy evolution models is therefore required to study and model the EBL\@. In practice, the EBL accumulates all photons emitted by radiative processes since the epoch of reionisation at $z \sim 6$, including direct stellar emission across ultraviolet, optical and near-infrared wavelengths, nebular line and continuum radiation, the contribution of active galactic nuclei (constrained observationally to $\lesssim 10\%$ at any relevant wavelength), and the thermally reprocessed component in which interstellar dust absorbs short-wavelength photons and re-radiates them in the mid- and 
far-infrared~\cite{10.1098/rsos.150555}. 
The latter component gives rise to the Cosmic Infrared Background (CIB), whose amplitude at $\lambda \gtrsim 8\,\mu\mathrm{m}$ is of comparable magnitude to the Cosmic Optical Background (COB) in the ultraviolet through near-infrared~\cite{annurev:/content/journals/10.1146/annurev.astro.39.1.249}. The relative contributions of these components evolve with redshift, broadly reflecting the cosmic star-formation rate density: at high redshift the UV background is comparatively stronger, while the infrared excess accumulated by dust reprocessing is lower, so that the EBL spectral energy distribution itself is a function of $z$ and must be treated as such in any line-of-sight integration of the optical depth (see Eq.~(\ref{eq:tau}) in the subsequent subsection).

Two complementary observational strategies are used to determine the EBL intensity. Deep galaxy number counts, obtained from UV to mid-infrared surveys with space observatories such as \textit{HST}, \textit{Spitzer} and \textit{Herschel}, integrate the light of individually resolved sources and provide strict lower limits on the EBL, since faint and intrinsically diffuse emission inevitably escapes detection~\cite{Hauser:2001xs}. Direct photometric measurements, which attempt to subtract all foreground contributions (most critically the zodiacal light from interplanetary dust in the Solar System, Galactic cirrus emission, and instrumental stray light), in principle capture the total background including any truly diffuse component~\cite{Hauser:2001xs}. However, the zodiacal light dominates over the EBL signal by one to two orders of magnitude at optical and near-infrared wavelengths, and its subtraction is a major systematic; as a consequence, direct measurements have persistently reported optical backgrounds exceeding the integrated galaxy counts by factors of two to three~\cite{Keenan_2010}, a tension that remains unresolved. The excess may reflect genuine diffuse emission from processes such as intra-halo light or tidally stripped stars, or alternatively may be attributable to systematic errors in the foreground model~\cite{doi:10.1126/science.1258168}.

The present uncertainties in the EBL spectral energy distribution (SED) are markedly wavelength-dependent. The COB is relatively well constrained in the optical and near-ultraviolet, while the near-infrared window at $1$--$5\,\mu\mathrm{m}$ carries both the largest absolute uncertainty and the greatest astrophysical importance for VHE $\gamma$-ray observations. This is a direct consequence of the kinematics of the pair-production process: the EBL photon energy most efficiently attenuating a $\gamma$-ray of energy $E_\gamma$ corresponds to the peak of the cross section (see Eq.\,(\ref{eq:sigma})), which maps to a target wavelength $\lambda_\mathrm{peak} \approx 1.24\,(E_\gamma/1\,\mathrm{TeV})\,\mu\mathrm{m}$. Consequently, IACT observations in the range $0.1$--$10\,\mathrm{TeV}$ are predominantly sensitive to the EBL in the window $0.1$--$12\,\mu\mathrm{m}$, precisely where the tension between direct photometry and galaxy counts is greatest and the infrared background most uncertain. The mid-infrared CIB above $\sim 8\,\mu\mathrm{m}$ is further compromised by the strong zodiacal foreground, while the far-infrared ($\sim 100\,\mu\mathrm{m}$) is comparatively better determined from \textit{Herschel} and \textit{Planck} observations.

An additional, speculative contribution to the optical and near-infrared background may arise from the radiative decay of massive axion-like particles (ALPs), light pseudo-scalar bosons predicted by extensions of the Standard Model, whose two-photon decay channel $a \to \gamma\gamma$ would produce a diffuse photon background potentially contributing to the excess observed in direct measurements. The distinctive spectral and redshift signatures expected from this mechanism have been constrained using blazar VHE spectra observed by \hess~\cite{Porras-Bedmar:2024uql}, setting upper limits on the ALP decay coupling that already inform viable models. The sensitivity of these constraints will improve substantially with the Cherenkov Telescope Array Observatory (CTAO), which will probe ALP contributions to the EBL at an order of magnitude lower optical depth than currently accessible.

Historically, large galaxy surveys carried out at UV, visible and infrared bands have been used to put lower limits on EBL levels, and galaxy counts have proved useful to construct semi-analytic models~\cite{Finke:2009xi,Gilmore:2011ks,2011MNRAS.410.2556D,Franceschini:2017iwq,2019MNRAS.484.4174K,2022ApJ...941...33F}. At the same time, VHE observations of $\gamma$-loud blazars served to constrain the EBL from above, establishing upper limits on the background intensity that disfavour the highest values from direct photometry~\cite{2006Natur.440.1018A}. Looking forward, the James Webb Space Telescope is detecting galaxy populations at flux levels previously inaccessible, systematically closing the gap between counts and direct photometry in the near-infrared and providing new constraints on the EBL out to $z \gtrsim 5$~\cite{Windhorst_2023}. Complementary $\gamma$-ray-based determinations of the EBL intensity itself have recently reached high significance~\cite{Banerjee:2026ebl,Baxter:2026tev}, providing an independent route to the tighter EBL normalisation on which the cosmological application presented here ultimately depends. Combined with the forthcoming VHE measurements from CTAO, which will extend the $\gamma$-ray opacity probe to higher redshifts and lower optical depths across the full IACT energy band, these developments are expected to substantially reduce the EBL systematic uncertainty that currently limits cosmological parameter inference from VHE blazar spectra.

\subsection{Attenuation of VHE \texorpdfstring{$\gamma$}{gamma} rays}
For a VHE photon travelling cosmological distances there is a non-zero probability of undergoing a photon-photon collision that leads to $e^+e^-$ pair production, the so-called Breit--Wheeler process~\cite{PhysRev.46.1087}. It is the main contribution to the cosmic $\gamma$-ray opacity and effectively prohibits any observations of VHE $\gamma$ rays from redshifts $z\gtrsim1$. The cross section for this process is given by:
\begin{equation}
    \sigma_{\gamma\gamma}(\beta) = \frac{3\sigma_T}{16}(1-\beta^2)\left[ (3-\beta^4) \ln\left(\frac{1+\beta}{1-\beta}\right) -2\beta(2-\beta^2) \right],
\label{eq:sigma}
\end{equation}
where $c\beta$ is the velocity of the electron in the centre-of-mass frame of the outgoing electrons. The threshold energy $\epsilon_{\rm th}$ for the interaction at an angle $\theta$ is:
\begin{equation}
    \epsilon_{\rm th} = \frac{2 (m_e c^2)^2}{E \mu (1+z)},
\end{equation}
where $\mu = 1- \cos\theta$, $\beta = \sqrt{1 - \epsilon_{\rm th}/\epsilon}$. $\epsilon$ and $E$ are the energy of the EBL and incident $\gamma$-ray photons, respectively.
The pair production cross section rises sharply above the energy threshold and is maximised for $E\epsilon =  4 m_e^2c^4$, which corresponds to an EBL photon energy of $\epsilon \simeq (1\,\rm TeV/E)$ eV, meaning that TeV photons interact most efficiently with infrared photons. 

\begin{figure}[!ht]
    \centering
    \includegraphics[width=0.8\linewidth]{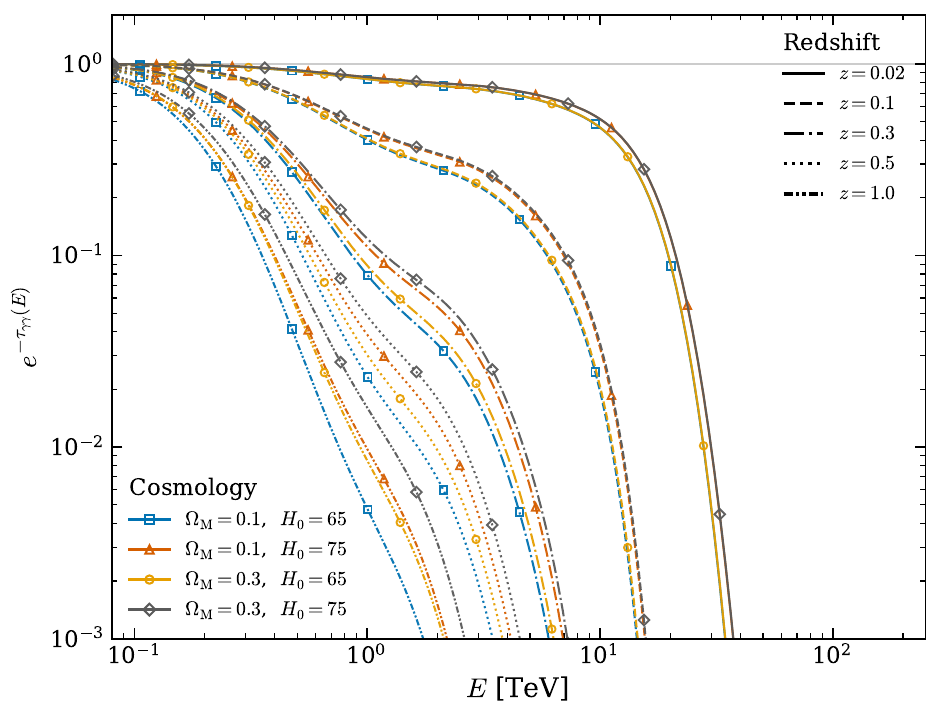}
    \caption{Absorption coefficient $e^{-\tau(E)}$ as a function of energy $E$ for different sets of cosmological parameters using the redshift-dependent EBL model from Ref.~\cite{2021MNRAS.507.5144S}. The absorption coefficient is computed for the source redshifts $z$ of 0.02 (solid line), 0.1 (dashed line), 0.3 (dotted-dashed line), 0.5 (dotted line) and 1 (double-dotted-dashed line), respectively, indicated by line type, and four sets of cosmology $(\Omega_\mathrm{M}, H_0)$ of $(0.1, 65)$, $(0.1, 75)$, $(0.3, 65)$ and $(0.3, 75)$, indicated by a colour and a marker. $H_0$ values are expressed in $\mathrm{km\,s^{-1}\,Mpc^{-1}}$.}
    \label{fig:EBL_cosmoparameters}
\end{figure}

The attenuation obtained via the optical depth $\tau_{\gamma\gamma}$ is expressed as:
\begin{equation}\label{eq:tau}
    \tau_{\gamma\gamma}(E, z_{\rm source}) = \int_0^{z_{\rm source}} \dd z \dv{l}{z}\int_0^2\dd\mu \frac{\mu}{2} \int_{\epsilon_{\rm th}}^{\infty} \dd\epsilon \frac{{\rm d} n_{\rm EBL}}{\rm d\epsilon}(\epsilon,z) \sigma_{\gamma\gamma}(E,\epsilon,\mu,z)\, ,
\end{equation}
where $\dd n_{\rm EBL}/\dd\epsilon$ is the specific number density of the EBL photons at a given energy $\epsilon$ and redshift $z$, as provided by the redshift-dependent EBL model.
For a given cosmology in a flat FLRW Universe, the line element is:
\begin{equation}
    \dv{l}{z} = c\left( H_0 (1+z)\sqrt{\Omega_\mathrm{M}(1+z)^3+\Omega_\Lambda} \right)^{-1},
\end{equation}
where $H_0$ is the Hubble constant, $\Omega_\mathrm{M}$ is the matter density parameter and $\Omega_\Lambda$ is the dark energy (cosmological constant) density parameter.
Throughout this work we also make use of the reduced Hubble constant defined as $h \equiv H_0/(100\,\mathrm{km\,s^{-1}\,Mpc^{-1}})$.
The cosmological parameters enter the optical depth not only through the line element above but also through the EBL photon field itself, whose construction from galaxy-survey data assumes luminosity distances and comoving volumes; however, since our sample (defined in section~\ref{sec:selection}) is contained within $z\lesssim1$, this dependence is weak and therefore can be neglected in the current approach.
Figure~\ref{fig:EBL_cosmoparameters} shows the absorption coefficient $\exp(-\tau(E))$ as a function of $\gamma$-ray energy $E$ for the baseline redshift-dependent EBL model~\cite{2021MNRAS.507.5144S} used in this work, source redshift $z$ values, and different sets of cosmological parameters ($H_0,\Omega_\mathrm{M}$). The absorption coefficient shows a clear dependence on these cosmological parameters and redshift. For $z = 0.3$, it varies by a factor of $\approx$ 1.5 at an energy of 1\,TeV between the extreme displayed sets of cosmological parameters, and by a factor of 2 near 1.5\,TeV.

\section{AGN selection for VHE \texorpdfstring{$\gamma$}{gamma}-ray observations}
\label{sec:selection}

\subsection{Sample from the high-energy \lat\ source catalogue}

The parent sample is drawn from the Third Catalog of Hard \textit{Fermi}-LAT Sources (3FHL)~\cite{Fermi-LAT:2017sxy}, which reports sources detected between $10\,\mathrm{GeV}$ and $2\,\mathrm{TeV}$ over seven years of survey data. This catalogue is particularly well suited to the present study: its hard spectral window directly adjoins the IACT energy range, so the catalogued spectra can be extrapolated to very high energies with a minimal lever arm, while the bulk of the photons that constrain each fit lies at energies low enough that the EBL optical depth is negligible ($\tau_{\gamma\gamma}\ll1$). The 3FHL fit therefore traces the \emph{intrinsic} source spectrum, which we extrapolate into the IACT band and subsequently attenuate by EBL absorption (see section~\ref{sec:analysis}) to predict the observed flux. This statement has a quantifiable limit: over the 20--200\,GeV range that dominates the catalogue fits, the residual attenuation corresponds to an effective index softening with a sample median of $\Delta\Gamma_\mathrm{eff} \simeq 0.05$ (growing with redshift, up to $\simeq 0.3$ for the most distant sources), although the photon-weighted impact on the fitted index is substantially smaller, since the fit statistics are dominated by the lowest energies where $\tau_{\gamma\gamma}$ is negligible. Because such an error would be coherent across the sample and correlated with redshift, we quantify its worst-case impact explicitly with the coherent-index mismodelling injections in section~\ref{sec:intrinsicmismodel}.

We select sources according to three requirements. First, we retain AGN of the blazar class, overwhelmingly BL\,Lac objects, which dominate the VHE extragalactic sky, whose hard GeV spectra make them viable targets for the photon-photon opacity probe. The smaller flat-spectrum radio quasar subset comprises 3C\,279, 4C\,+21.35, PKS\,1441+25, PKS\,1510$-$089 and the gravitationally lensed S3\,0218+35. For these objects the extrapolation from the GeV-band anchor is the most delicate, and the source-level hierarchy of section~\ref{sec:analysis} is designed to accommodate such departures. Second, we require a secure measured spectroscopic redshift, since the source distance enters the optical-depth computation (see Eq.~(\ref{eq:tau})) and is indispensable for predicting the EBL attenuation. Third, we require at least one published VHE spectrum in the STeVECat compilation from \hess, MAGIC or VERITAS observations (see section~\ref{sec:observability}), passing the quality cuts described there. Applying these criteria yields a sample of 50 AGN spanning the redshift range $z = 0.018$--$0.944$ (median source redshift $z \simeq 0.13$), listed together with their coordinates and redshifts in table~\ref{tab:agn_sample}. For each source we adopt the best-fit spectral model reported in the 3FHL catalogue as the anchor for the intrinsic spectrum. Two parametrisations occur: a power law (PL) given by
\begin{equation}
    \frac{\mathrm{d}\Phi^{\rm int}_\gamma}{\mathrm{d}E}
        = \Phi_0 \left(\frac{E}{E_0}\right)^{-\Gamma},
\end{equation}
adopted for 44 of the sources, and a log-parabola (LP),
\begin{equation}
    \frac{\mathrm{d}\Phi^{\rm int}_\gamma}{\mathrm{d}E}
        = \Phi_0 \left(\frac{E}{E_0}\right)^{-\left(\alpha + \beta\,\ln(E/E_0)\right)},
\end{equation}
adopted for the remaining 6, for which the log-parabola parametrisation is significantly preferred over the power-law one in the catalogue. No artificial intrinsic spectral cutoff is imposed on either template. The corresponding best-fit parameters are collected in table~\ref{tab:agn_sample2}: the flux normalisation $\Phi_0$, evaluated at a reference energy of $100\,\mathrm{GeV}$, and the spectral shape parameters $(\Gamma)$ or $(\alpha,\beta)$. For reference we also quote in the same table the photon index $\Gamma_{\rm 3FHL}$ of a pure power-law fit to the 3FHL data, as a uniform reference measure of the spectral hardness of each source; given the residual attenuation discussed above, it is not assumed to be an exact intrinsic index. These parametrised spectra constitute a proxy for the intrinsic emission $\mathrm{d}\Phi^{\rm int}_\gamma/\mathrm{d}E$ that, once attenuated by the energy- and redshift-dependent factor $e^{-\tau_{\gamma\gamma}(E,z)}$, defines the observable flux modelled in section~\ref{sec:analysis}. Their catalogue uncertainties and possible GeV-to-TeV spectral evolution are propagated by the source-level hierarchy introduced there.
\begin{table}[ht!]
\footnotesize
    \centering
    \begin{tabular}{l|c|c|c|l}
\hline\hline
\lat Name & RA & Dec & $z$ & Common Name \\ \hline\hline
3FHL J0319.8+1845 & 3 19 51.81 & +18 45 34.56 & 0.19 & 1E 0317.0+1835 \\
3FHL J0232.8+2017 & 2 32 48.61 & +20 17 17.49 & 0.139 & 1ES 0229+200 \\
3FHL J0349.3-1159 & 3 49 23.19 & -11 59 27.37 & 0.188 & 1ES 0347-121 \\
3FHL J0416.8+0105 & 4 16 52.49 & +1 5 23.90 & 0.287 & 1ES 0414+009 \\
3FHL J0809.8+5218 & 8 9 49.19 & +52 18 58.25 & 0.138 & 1ES 0806+524 \\
3FHL J1103.6-2328 & 11 3 37.61 & -23 29 31.20 & 0.186 & 1ES 1101-232 \\
3FHL J1442.8+1200 & 14 42 48.28 & +12 0 40.35 & 0.163 & 1ES 1440+122 \\
3FHL J1959.9+6508 & 19 59 59.85 & +65 8 54.65 & 0.047 & 1ES 1959+650 \\
3FHL J2347.0+5142 & 23 47 4.84 & +51 42 17.88 & 0.044 & 1ES 2344+514 \\
3FHL J0710.4+5908 & 7 10 30.07 & +59 8 20.37 & 0.125 & 1H 0658+595 \\
3FHL J1015.0+4926 & 10 15 4.14 & +49 26 0.71 & 0.212 & 1H 1013+498 \\
3FHL J1010.2-3119 & 10 10 15.98 & -31 19 8.41 & 0.142639 & 1RXS J101015.9-311909 \\
3FHL J1145.0+1935 & 11 45 5.01 & +19 36 22.74 & 0.0216 & 3C 264 \\
3FHL J1256.1-0547 & 12 56 11.17 & -5 47 21.53 & 0.536 & 3C 279 \\
3FHL J1224.9+2122 & 12 24 54.46 & +21 22 46.38 & 0.434 & 4C +21.35 \\
3FHL J1517.6-2422 & 15 17 41.81 & -24 22 19.48 & 0.048 & AP Librae \\
3FHL J1217.9+3006 & 12 17 52.08 & +30 7 0.63 & 0.13 & B2 1215+30 \\
3FHL J2250.0+3825 & 22 50 5.75 & +38 24 37.19 & 0.119 & B3 2247+381 \\
3FHL J2202.7+4216 & 22 2 43.29 & +42 16 39.98 & 0.069 & BL Lacertae \\
3FHL J1428.5+4240 & 14 28 32.61 & +42 40 21.05 & 0.129 & H 1426+428 \\
3FHL J2359.1-3038 & 23 59 7.90 & -30 37 40.67 & 0.165 & H 2356-309 \\
3FHL J1728.3+5013 & 17 28 18.62 & +50 13 10.47 & 0.055 & I Zw 187 \\
3FHL J0316.6+4120 & 3 16 42.98 & +41 19 29.62 & 0.01894 & IC 310 \\
3FHL J2042.0+2428 & 20 42 6.05 & +24 26 52.34 & 0.104 & MG2 J204208+2426 \\
3FHL J2001.2+4353 & 20 1 13.50 & +43 53 2.80 & 0.174 & MG4 J200112+4352 \\
3FHL J1315.0-4237 & 13 15 3.39 & -42 36 49.76 & 0.105 & MS 13121-4221 \\
3FHL J1136.5+7009 & 11 36 26.41 & +70 9 27.31 & 0.045 & Mkn 180 \\
3FHL J1104.4+3812 & 11 4 27.31 & +38 12 31.80 & 0.03 & Mkn 421 \\
3FHL J1653.8+3945 & 16 53 52.22 & +39 45 36.61 & 0.033 & Mkn 501 \\
3FHL J0319.8+4130 & 3 19 48.16 & +41 30 42.11 & 0.018 & NGC 1275 \\
3FHL J0733.4+5152 & 7 33 26.79 & +51 53 55.91 & 0.065 & NVSS J073326+515355 \\
3FHL J1221.3+3010 & 12 21 21.94 & +30 10 37.16 & 0.184 & PG 1218+304 \\
3FHL J0303.4-2407 & 3 3 26.50 & -24 7 11.43 & 0.266 & PKS 0301-243 \\
3FHL J0627.1-3528 & 6 27 6.73 & -35 29 15.36 & 0.05494 & PKS 0625-354 \\
3FHL J1443.9-3908 & 14 43 57.20 & -39 8 40.05 & 0.1385 & PKS 1440-389 \\
3FHL J1443.9+2502 & 14 43 56.89 & +25 1 44.49 & 0.939 & PKS 1441+25 \\
3FHL J1512.8-0906 & 15 12 50.53 & -9 5 59.83 & 0.36 & PKS 1510-089 \\
3FHL J2009.4-4849 & 20 9 25.39 & -48 49 53.73 & 0.071 & PKS 2005-489 \\
3FHL J2158.8-3013 & 21 58 52.07 & -30 13 32.12 & 0.116 & PKS 2155-304 \\
3FHL J0152.6+0147 & 1 52 39.61 & +1 47 17.38 & 0.08 & PMN J0152+0146 \\
3FHL J0648.7+1517 & 6 48 47.65 & +15 16 24.80 & 0.179 & RX J0648.7+1516 \\
3FHL J0847.2+1134 & 8 47 12.93 & +11 33 50.15 & 0.198199 & RX J0847.1+1133 \\
3FHL J0112.1+2245 & 1 12 5.82 & +22 44 38.79 & 0.265 & S2 0109+22 \\
3FHL J0221.1+3556 & 2 21 5.49 & +35 56 13.85 & 0.944 & S3 0218+35 \\
3FHL J1744.0+1935 & 17 43 57.83 & +19 35 9.02 & 0.084 & S3 1741+19 \\
3FHL J0013.8-1855 & 0 13 56.04 & -18 54 6.70 & 0.095 & SHBL J001355.9-185406 \\
3FHL J0214.5+5145 & 2 14 17.94 & +51 44 51.95 & 0.049 & TXS 0210+515 \\
3FHL J0509.4+0542 & 5 9 25.96 & +5 41 35.33 & 0.3365 & TXS 0506+056 \\
3FHL J1518.0-2731 & 15 18 3.60 & -27 31 31.19 & 0.1285 & TXS 1515-273 \\
3FHL J1221.5+2813 & 12 21 31.69 & +28 13 58.50 & 0.102 & W Comae \\
\hline\hline
\end{tabular}
    \caption{The 50 AGN entering the analysis sample, selected by cross-matching the STeVECat compilation of published VHE spectra with the \texttt{3FHL} catalogue~\cite{Fermi-LAT:2017sxy} (section~\ref{sec:selection}). Columns: 3FHL designation, J2000 equatorial coordinates (h m s, $^{\circ}$ $'$ $''$), spectroscopic redshift $z$ (STeVECat, secure-flag only), and the conventional source name. The sample spans $z = 0.018$--$0.944$ with median $z \simeq 0.129$.}
    \label{tab:agn_sample}
\end{table}

\begin{table}[ht!]
\scriptsize
    \centering
    \begin{tabular}{l|c|c|c}
\hline\hline
3FHL source & $\Gamma_{\rm 3FHL}$ & Parametrisation & Best-fit parameters \\ \hline\hline
3FHL J0319.8+1845 & 1.89 & PL & $(124,\ 1.89)$ \\
3FHL J0232.8+2017 & 1.81 & PL & $(62.6,\ 1.81)$ \\
3FHL J0349.3-1159 & 1.65 & PL & $(89,\ 1.65)$ \\
3FHL J0416.8+0105 & 1.98 & PL & $(107,\ 1.98)$ \\
3FHL J0809.8+5218 & 2.23 & PL & $(366,\ 2.23)$ \\
3FHL J1103.6-2328 & 1.67 & PL & $(121,\ 1.67)$ \\
3FHL J1442.8+1200 & 2.05 & PL & $(69.7,\ 2.05)$ \\
3FHL J1959.9+6508 & 1.94 & PL & $(916,\ 1.94)$ \\
3FHL J2347.0+5142 & 1.85 & PL & $(408,\ 1.85)$ \\
3FHL J0710.4+5908 & 1.91 & LP & $(129,\ 2.93,\ 0.639)$ \\
3FHL J1015.0+4926 & 2.15 & PL & $(884,\ 2.15)$ \\
3FHL J1010.2-3119 & 1.65 & PL & $(167,\ 1.65)$ \\
3FHL J1145.0+1935 & 1.65 & PL & $(37.7,\ 1.65)$ \\
3FHL J1256.1-0547 & 2.86 & PL & $(140,\ 2.86)$ \\
3FHL J1224.9+2122 & 2.63 & PL & $(259,\ 2.63)$ \\
3FHL J1517.6-2422 & 2.26 & PL & $(259,\ 2.26)$ \\
3FHL J1217.9+3006 & 2.29 & PL & $(501,\ 2.29)$ \\
3FHL J2250.0+3825 & 1.65 & PL & $(195,\ 1.65)$ \\
3FHL J2202.7+4216 & 2.64 & PL & $(218,\ 2.64)$ \\
3FHL J1428.5+4240 & 1.92 & PL & $(199,\ 1.92)$ \\
3FHL J2359.1-3038 & 2.29 & PL & $(53.9,\ 2.29)$ \\
3FHL J1728.3+5013 & 2.00 & PL & $(296,\ 2.00)$ \\
3FHL J0316.6+4120 & 1.49 & PL & $(75.4,\ 1.49)$ \\
3FHL J2042.0+2428 & 1.88 & PL & $(67.1,\ 1.88)$ \\
3FHL J2001.2+4353 & 2.47 & PL & $(260,\ 2.47)$ \\
3FHL J1315.0-4237 & 1.87 & PL & $(88.4,\ 1.87)$ \\
3FHL J1136.5+7009 & 1.88 & PL & $(204,\ 1.88)$ \\
3FHL J1104.4+3812 & 1.95 & LP & $(6.92e+03,\ 2.07,\ 0.096)$ \\
3FHL J1653.8+3945 & 1.86 & LP & $(2.59e+03,\ 2.01,\ 0.125)$ \\
3FHL J0319.8+4130 & 2.87 & PL & $(359,\ 2.87)$ \\
3FHL J0733.4+5152 & 1.76 & PL & $(47.2,\ 1.76)$ \\
3FHL J1221.3+3010 & 1.86 & LP & $(825,\ 2.13,\ 0.217)$ \\
3FHL J0303.4-2407 & 2.20 & PL & $(482,\ 2.20)$ \\
3FHL J0627.1-3528 & 2.06 & PL & $(167,\ 2.06)$ \\
3FHL J1443.9-3908 & 2.15 & PL & $(453,\ 2.15)$ \\
3FHL J1443.9+2502 & 2.86 & PL & $(51.7,\ 2.86)$ \\
3FHL J1512.8-0906 & 2.73 & PL & $(307,\ 2.73)$ \\
3FHL J2009.4-4849 & 1.96 & PL & $(413,\ 1.96)$ \\
3FHL J2158.8-3013 & 2.14 & LP & $(2.06e+03,\ 2.39,\ 0.150)$ \\
3FHL J0152.6+0147 & 2.11 & PL & $(108,\ 2.11)$ \\
3FHL J0648.7+1517 & 1.83 & PL & $(336,\ 1.83)$ \\
3FHL J0847.2+1134 & 1.86 & LP & $(92.9,\ 3.56,\ 0.988)$ \\
3FHL J0112.1+2245 & 2.90 & PL & $(124,\ 2.90)$ \\
3FHL J0221.1+3556 & 2.54 & PL & $(125,\ 2.54)$ \\
3FHL J1744.0+1935 & 2.20 & PL & $(61.3,\ 2.20)$ \\
3FHL J0013.8-1855 & 2.60 & PL & $(12.5,\ 2.60)$ \\
3FHL J0214.5+5145 & 1.55 & PL & $(71.1,\ 1.55)$ \\
3FHL J0509.4+0542 & 2.33 & PL & $(248,\ 2.33)$ \\
3FHL J1518.0-2731 & 2.17 & PL & $(85.7,\ 2.17)$ \\
3FHL J1221.5+2813 & 2.26 & PL & $(138,\ 2.26)$ \\
\hline\hline
\end{tabular}
    \caption{3FHL spectral anchors for the sample~\cite{Fermi-LAT:2017sxy}: 44 power laws (PL) and 6 log-parabolas (LP). The second column gives the photon index $\Gamma_{\rm 3FHL}$ of a pure power-law fit as a uniform reference measure of spectral hardness; it is not assumed to be an exact intrinsic index. The last column gives $(\Phi_0,\Gamma)$ for PL and $(\Phi_0,\alpha,\beta)$ for LP anchors. The normalisation $\Phi_0$, in $10^{-12}\,\mathrm{cm^{-2}\,s^{-1}\,TeV^{-1}}$, and LP index $\alpha$ refer to $E_0=100$\,GeV, with $\alpha(E_0)=\alpha(E_{\rm piv})+2\beta\ln(E_0/E_{\rm piv})$ in the natural-log convention. Source-specific catalogue uncertainties in index and curvature enter the hierarchy but are omitted here for compactness.}
    \label{tab:agn_sample2}
\end{table}

\subsection{Observations with imaging atmospheric Cherenkov telescopes}
\label{sec:observability}
Aiming to evaluate the sensitivity of IACTs to cosmological parameters via spectral measurements of the energy-differential flux of $\gamma$ rays from selected AGN, we further restrict the selected AGN sample to objects actually observed by ongoing IACTs. In what follows, we will therefore consider archival and mock observations by \hess, MAGIC and VERITAS in the VHE band.

The \hess\ observatory is composed of an array of five IACTs, four of 12\,m diameter and one of 28\,m diameter, situated in the Khomas Highlands of Namibia, 23$^\circ$16$'$18$''$S, 16$^\circ$30$'$0$''$E, at 1.8\,km a.s.l.~\cite{HESS}. More than 1300 hours of data taking per year are obtained. Representative energy-dependent instrument response functions (IRFs) for \hess\ observations at a zenith angle of 20$^\circ$ are extracted from Ref.~\cite{HESS:2022ygk}.
These IRFs are obtained from an average over a large amount of observations taken at different zenith angles and for various instrumental and atmospheric conditions. MAGIC is an IACT array of two 17\,m-diameter telescopes located 2.2\,km a.s.l., 28$^\circ$45$'$43$''$N, 17$^\circ$53$'$24$''$W, at the Roque de los Muchachos Observatory, on the Canary Island of La Palma, Spain~\cite{MAGIC}. Representative energy-dependent IRFs for MAGIC observations at a 20$^\circ$ zenith angle are based on Ref.~\cite{Abe:2023kvd}.
The four-IACT array VERITAS is located in southern Arizona at 31$^\circ$40$'$30$''$N, 110$^\circ$57$'$07$''$W, 1.3 km a.s.l.~\cite{VERITAS}.  A representative set of energy-dependent IRFs for VERITAS observations under 20$^\circ$ zenith angle is taken from Ref.~\cite{Adams:2021hzq}. Publicly released, analysis-grade IRFs are not available for these observatories. The representative response curves therefore enter only the counts-level forecasting machinery of section~\ref{sec:analysis}, while all results in this work are derived from the published flux points, which are already corrected for the instrument response by the original publications.

The VHE observations themselves are taken from STeVECat~\cite{Greaux:2023stevecat}, a homogenised compilation of all published spectra from the major IACTs. A \emph{dataset} in this work corresponds to one published spectrum of one source by one 
instrument over one observation campaign. We retain only the non-overlapping subset of STeVECat (\texttt{overlap\_flag} $= 1$), in which no two spectra of the same source share observation epochs, so that all datasets are statistically independent observations. A source observed by several instruments, or by the same instrument in different campaigns, contributes correspondingly to many datasets. Each dataset is further required to contain at least three detected flux points, the minimal number that still constrains a normalisation and a residual shape distortion within one dataset. For the 50 selected AGN this yields 241 independent datasets in total: 40 by \hess\ (from 19 sources), 140 by MAGIC and 61 by VERITAS, with per-source counts, summed livetimes and the original publications listed in table~\ref{tab:agn_sample_iacts}. The published flux points span approximately $0.05$--$20$\,TeV depending on the source and instrument. All results in this work are derived from the full three-array sample of 50 sources, comprising 241 independent datasets with 1922 detected flux points in total. 
\begin{table}[ht!]
\scriptsize
    \centering
    \resizebox{\textwidth}{!}{%
    \begin{tabular}{l|r r|r r|r r}
\hline\hline
 & \multicolumn{2}{c|}{H.E.S.S.} & \multicolumn{2}{c|}{MAGIC} & \multicolumn{2}{c}{VERITAS} \\
\lat Name & $N_{\rm ds}$ ($T_{\rm obs}$ [h]) & Ref. & $N_{\rm ds}$ ($T_{\rm obs}$ [h]) & Ref. & $N_{\rm ds}$ ($T_{\rm obs}$ [h]) & Ref. \\ \hline\hline
3FHL J0319.8+1845 & -- & -- & -- & -- & 1 (26.0) & \cite{2012ApJ...750...94A} \\
3FHL J0232.8+2017 & 1 (72.3) & \cite{2014AA...562A.145H} & 1 (105.2) & \cite{2019MNRAS.486.4233A} & 2 (--) & \cite{2014ApJ...782...13A} \\
3FHL J0349.3-1159 & 1 (25.4) & \cite{2007AA...473L..25A} & -- & -- & -- & -- \\
3FHL J0416.8+0105 & 1 (73.7) & \cite{2012AA...538A.103H} & -- & -- & 1 (56.2) & \cite{2012ApJ...755..118A} \\
3FHL J0809.8+5218 & -- & -- & 2 (16.1) & \cite{2015MNRAS.451..739A} & 1 (35.0) & \cite{2009ApJ...690L.126A} \\
3FHL J1103.6-2328 & 1 (62.9) & \cite{2014AA...562A.145H} & -- & -- & -- & -- \\
3FHL J1442.8+1200 & -- & -- & -- & -- & 1 (53.0) & \cite{2016MNRAS.461..202A} \\
3FHL J1959.9+6508 & -- & -- & 8 (27.7) & \cite{2006ApJ...639..761A, 2008ApJ...679.1029T, 2019MNRAS.486.4233A, 2020AA...638A..14M, 2020AA...640A.132M} & 3 (16.3) & \cite{2013ApJ...775....3A, 2014ApJ...797...89A} \\
3FHL J2347.0+5142 & -- & -- & 3 (44.4) & \cite{2007ApJ...662..892A, 2013AA...556A..67A, 2020AA...640A.132M} & 2 (47.2) & \cite{2011ApJ...738..169A, 2017MNRAS.471.2117A} \\
3FHL J0710.4+5908 & -- & -- & -- & -- & 1 (22.1) & \cite{2010ApJ...715L..49A} \\
3FHL J1015.0+4926 & -- & -- & 4 (80.5) & \cite{2007ApJ...667L..21A, 2016AA...591A..10A, 2016MNRAS.459.2286A, 2019MNRAS.486.4233A} & -- & -- \\
3FHL J1010.2-3119 & 1 (48.7) & \cite{2012AA...542A..94H} & -- & -- & -- & -- \\
3FHL J1145.0+1935 & -- & -- & -- & -- & 1 (37.9) & \cite{2020ApJ...896...41A} \\
3FHL J1256.1-0547 & 1 (2.2) & \cite{2019AA...627A.159H} & 2 (20.0) & \cite{2008Sci...320.1752M, 2011AA...530A...4A} & -- & -- \\
3FHL J1224.9+2122 & -- & -- & 1 (0.5) & \cite{2011ApJ...730L...8A} & -- & -- \\
3FHL J1517.6-2422 & 1 (14.0) & \cite{2015AA...573A..31H} & -- & -- & -- & -- \\
3FHL J1217.9+3006 & -- & -- & 1 (20.3) & \cite{2012AA...544A.142A} & 9 (160.2) & \cite{2017ApJ...836..205A, 2020ApJ...891..170V} \\
3FHL J2250.0+3825 & -- & -- & 1 (14.2) & \cite{2012AA...539A.118A} & -- & -- \\
3FHL J2202.7+4216 & -- & -- & 3 (27.4) & \cite{2007ApJ...666L..17A, 2019AA...623A.175M} & 2 (2.9) & \cite{2013ApJ...762...92A, 2018ApJ...856...95A} \\
3FHL J1428.5+4240 & -- & -- & 1 (8.7) & \cite{2020ApJS..247...16A} & -- & -- \\
3FHL J2359.1-3038 & 3 (109.8) & \cite{2010AA...516A..56H} & -- & -- & -- & -- \\
3FHL J1728.3+5013 & -- & -- & 1 (6.4) & \cite{2019MNRAS.486.4233A} & 1 (6.0) & \cite{2015ApJ...808..110A} \\
3FHL J0316.6+4120 & -- & -- & 8 (82.2) & \cite{2014AA...563A..91A, 2017AA...603A..25A} & -- & -- \\
3FHL J2042.0+2428 & -- & -- & 1 (52.5) & \cite{2020ApJS..247...16A} & -- & -- \\
3FHL J2001.2+4353 & -- & -- & 1 (1.4) & \cite{2014AA...572A.121A} & -- & -- \\
3FHL J1315.0-4237 & 1 (48.4) & \cite{2013MNRAS.434.1889H} & -- & -- & -- & -- \\
3FHL J1136.5+7009 & -- & -- & 1 (11.1) & \cite{2006ApJ...648L.105A} & -- & -- \\
3FHL J1104.4+3812 & 1 (14.7) & \cite{2005AA...437...95A} & 37 (203.8) & \cite{2007ApJ...663..125A, 2010AA...519A..32A, 2011ApJ...736..131A, 2012AA...542A.100A, 2016ApJ...819..156B, 2019MNRAS.486.4233A, 2021AA...655A..89M} & 12 (49.6) & \cite{2011ApJ...738...25A, 2016ApJ...819..156B, 2017ApJ...834....2A} \\
3FHL J1653.8+3945 & 1 (1.8) & \cite{2019ApJ...870...93A} & 39 (63.0) & \cite{2007ApJ...669..862A, 2009ApJ...705.1624A, 2015AA...573A..50A, 2015ApJ...812...65F, 2017AA...603A..31A, 2018AA...620A.181A, 2020AA...637A..86M} & 16 (10.8) & \cite{2011ApJ...727..129A, 2015AA...573A..50A, 2015ApJ...812...65F, 2016AA...594A..76A, 2018AA...620A.181A} \\
3FHL J0319.8+4130 & -- & -- & 4 (257.1) & \cite{2016AA...589A..33A, 2018AA...617A..91M} & -- & -- \\
3FHL J0733.4+5152 & -- & -- & 1 (23.4) & \cite{2019MNRAS.490.2284M} & -- & -- \\
3FHL J1221.3+3010 & -- & -- & 1 (8.2) & \cite{2006ApJ...642L.119A} & 2 (44.6) & \cite{2009ApJ...695.1370A, 2010ApJ...709L.163A} \\
3FHL J0303.4-2407 & 1 (34.9) & \cite{2013AA...559A.136H} & -- & -- & -- & -- \\
3FHL J0627.1-3528 & 1 (5.5) & \cite{2018MNRAS.476.4187H} & -- & -- & -- & -- \\
3FHL J1443.9-3908 & 1 (11.9) & \cite{2020MNRAS.494.5590A} & -- & -- & -- & -- \\
3FHL J1443.9+2502 & -- & -- & 2 (--) & \cite{2015ApJ...815L..23A} & 1 (15.0) & \cite{2015ApJ...815L..22A} \\
3FHL J1512.8-0906 & 2 (17.6) & \cite{2013AA...554A.107H, 2021AA...648A..23H} & 7 (102.2) & \cite{2014AA...569A..46A, 2017AA...603A..29A, 2018AA...619A.159M, 2019MNRAS.486.4233A, 2021AA...648A..23H} & -- & -- \\
3FHL J2009.4-4849 & 5 (103.3) & \cite{2010AA...511A..52H, 2011AA...533A.110H} & -- & -- & -- & -- \\
3FHL J2158.8-3013 & 15 (316.5) & \cite{2005AA...430..865A, 2005AA...442..895A, 2012AA...539A.149H, 2014AA...562A.145H, 2017AA...600A..89H} & 1 (15.0) & \cite{2012AA...544A..75A} & -- & -- \\
3FHL J0152.6+0147 & 1 (14.7) & \cite{2008AA...481L.103A} & -- & -- & -- & -- \\
3FHL J0648.7+1517 & -- & -- & -- & -- & 1 (19.3) & \cite{2011ApJ...742..127A} \\
3FHL J0847.2+1134 & -- & -- & 1 (45.3) & \cite{2020ApJS..247...16A} & -- & -- \\
3FHL J0112.1+2245 & -- & -- & 1 (1.4) & \cite{2018MNRAS.480..879M} & -- & -- \\
3FHL J0221.1+3556 & -- & -- & 1 (2.1) & \cite{2016AA...595A..98A} & -- & -- \\
3FHL J1744.0+1935 & -- & -- & 1 (46.0) & \cite{2017MNRAS.468.1534A} & 1 (30.0) & \cite{2016MNRAS.459.2550A} \\
3FHL J0013.8-1855 & 1 (41.5) & \cite{2013AA...554A..72H} & -- & -- & -- & -- \\
3FHL J0214.5+5145 & -- & -- & 1 (28.6) & \cite{2020ApJS..247...16A} & -- & -- \\
3FHL J0509.4+0542 & -- & -- & 3 (39.7) & \cite{2018ApJ...863L..10A} & 1 (34.9) & \cite{2018ApJ...861L..20A} \\
3FHL J1518.0-2731 & -- & -- & 1 (8.1) & \cite{2021MNRAS.507.1528A} & -- & -- \\
3FHL J1221.5+2813 & -- & -- & -- & -- & 2 (43.3) & \cite{2008ApJ...684L..73A, 2009ApJ...707..612A} \\
\hline\hline
    \end{tabular}%
    }
    \caption{Archival VHE datasets entering the analysis, from the STeVECat compilation: for each source and instrument, the number of independent datasets $N_{\rm ds}$ (non-overlapping observation periods, STeVECat \texttt{overlap\_flag} $= 1$) with the summed livetime in hours in parentheses (`--' where no livetime is reported), and the references of the original publications. In total 241 datasets: 40 by H.E.S.S., 140 by MAGIC and 61 by VERITAS, respectively. Each dataset is treated as an independent observation in the joint likelihood analysis.}
    \label{tab:agn_sample_iacts}
\end{table}

\section{Data analysis of combined observations of selected AGN with IACTs}
\label{sec:analysis}
\subsection{Expected \texorpdfstring{$\gamma$}{gamma}-ray signals}
The observed energy-differential flux for a source $s$ is given by:
\begin{equation}
\frac{d\Phi_{\rm obs, s}(E)}{dE}= \frac{d\Phi_{\rm int, s}(E)}{dE} \times  e^{-\tau(E,z_s)}\, ,
    \label{eq:flux}
\end{equation}
where $d\Phi_{\rm int, s}/dE$ is the intrinsic flux of the source $s$, anchored to its 3FHL power-law or log-parabola parametrisation of section~\ref{sec:selection}. Its index and curvature are allowed to depart from the catalogue values through the source-level hierarchy introduced below, and no additional intrinsic cutoff is imposed. The observed spectrum is attenuated by the absorption coefficient $e^{-\tau(E,z)}$.
The model-expected number of events $ N^\mathcal{M}_{ik}$ in the detector for the $k$-th dataset, of parent source $s(k)$, in the $i$-th energy bin of width $\Delta E_i$ can be written as:
\begin{equation}
\label{eq:count}
N^\mathcal{M}_{ik} = T_{{\rm obs}, k} \int_{E_i - \Delta E_i /2}^{E_i + \Delta E_i /2} dE  \int_{-\infty}^{\infty} dE'\, \frac{d\Phi_{{\rm obs}, s(k)}(E')}{dE'}\,
    A_{{\rm eff}, k}^{\gamma}(E')\, G(E - E')\,.
\end{equation}
The finite energy resolution of the instrument $G$ is modelled as a Gaussian with width given by $\sigma/E$, with $E'$ and $E$ the true and reconstructed energies, respectively. $T_{\rm obs,k}$ corresponds to the observation
time of the dataset $k$, and $A_{\rm eff, k}$
is the energy-dependent effective area to VHE
$\gamma$ rays for a given dataset $k$ due to the zenith-angle dependence of $A_{\rm eff, k}$. When mock counts are generated, they are binned from $0.1$ to $30$\,TeV using 10 logarithmically spaced spectral bins per energy decade. For the archival analysis of the 241 datasets discussed in section~\ref{sec:selection}, the observables are instead the published flux points at their original energies, and the Gaussian flux-point likelihood of
Eq.~(\ref{eq:gausslik}) replaces the counts-based formalism; the mock flux points used in the injection studies are generated at the same energies with each dataset's measured fractional uncertainties, which preserves the realistic per-instrument sensitivity profile without requiring effective-area assumptions.

Working with published flux points involves a number of assumptions. The points are treated as independent Gaussian measurements: asymmetric published uncertainties are symmetrised, correlations between neighbouring points introduced by the original unfolding of the instrument response are neglected, and no correlated inter-dataset systematic (such as a common energy-scale uncertainty of $\sim$10--15\% per instrument, which can partially mimic an opacity change) is modelled. These simplifications are imposed by the archival nature of the data: analysis-grade IRFs and event-level counts are not publicly available for the bulk of the sample, and adopting approximate response curves would turn the analysis into a simulation rather than a measurement. A counts-level joint analysis with official IRFs remains the natural upgrade path, and the counts formalism above is retained for that purpose and for the mock forecasting. Within these limitations, the free per-dataset normalisation introduced below removes the leading (achromatic) inter-epoch systematic uncertainty, and the shape-level systematic uncertainties are quantified explicitly in sections~\ref{sec:eblmismodel} and~\ref{sec:intrinsicmismodel}.

\subsection{Statistical data analysis}
The statistical analysis is cast as a Bayesian inference problem. The target quantity is the joint posterior distribution of the opacity ratio $r\equiv\alpha_{\rm EBL}/h$, defined through Eq.~(\ref{eq:rdef}) below, and the matter density $\Omega_\mathrm{M}$, together with a set of hyperparameters describing departures of the intrinsic VHE spectra from their 3FHL anchors. 

In the hierarchical model, each source is assigned one correction to its 3FHL spectral index and one intrinsic-curvature parameter, and the same pair is used for every observation of that source. The source-specific parameters are assigned common population distributions, whose means and dispersions are inferred jointly with the cosmological parameters. Multiple observations of the same object therefore constrain the same latent spectral shape rather than entering the population as independent sources. By contrast, the flux normalisation remains specific to each dataset and accounts for changes in the activity state between observing campaigns. This construction allows for both coherent and source-to-source departures from the 3FHL spectra, while using information from the full sample to constrain source parameters that are only weakly determined by an individual spectrum.

Hierarchical (i.e., multilevel) modelling of this kind is an established tool of astronomical population and cosmological inference~\cite{Gelman:2013bda,2007ApJ...665.1489K}, underpinning modern Type~Ia supernova and distance-ladder determinations of $H_0$~\cite{2011MNRAS.418.2308M,2018MNRAS.476.3861F} as well as gravitational-wave population analyses~\cite{2019PASA...36...10T}. In $\gamma$-ray astronomy it has so far appeared in population-level treatments of blazar spectral-index distributions~\cite{2007ApJ...666..128V} and, recently, in hierarchical Bayesian tests of Lorentz invariance with $\gamma$-ray bursts~\cite{2025arXiv251222875D}, but, to our knowledge, it has not previously been applied to the $\gamma$-ray opacity probe of cosmology. The $\gamma$-ray likelihood does not separately identify $h$ and $\alpha_{\rm EBL}$ as mentioned above; $H_0$ is consequently mostly derived only after an external assumption on the EBL scale has been declared.

\subsubsection{Likelihood function and source hierarchy}
\label{sec:likelihood}
The observed counts $n_{ik}$ in energy bin $i$ for dataset $k$ are Poisson-distributed with mean $N_{ik}^{\mathcal{M}}$ given by Eq.\,(\ref{eq:count}). The Poisson log-likelihood can therefore be written as:
\begin{equation} \label{eq:loglik}
\ln \mathcal{L} = \sum_{k}\sum_{i} \left[ n_{ik}\ln N_{ik}^{\mathcal{M}} - N_{ik}^{\mathcal{M}} \right]\, ,
\end{equation}
omitting the $-\ln(n_{ik}!)$ term which is constant with respect to the model parameters. The equivalent Cash statistic \citep{Cash1979}, given by:
\begin{equation} \label{eq:cash}
C = -2\ln\mathcal{L} = 2\sum_{k}\sum_{i} \left[ N_{ik}^{\mathcal{M}} - n_{ik} + n_{ik}\ln\!\frac{n_{ik}}{N_{ik}^{\mathcal{M}}} \right]\, ,
\end{equation}
applies to this event-level analysis mode, retained as the natural upgrade path discussed above; it reduces to $\chi^2$ in the large-count limit. The total log-likelihood of the combined dataset is the sum $\ln\mathcal{L}_\mathrm{tot} = \sum_k \ln\mathcal{L}_k$, exploiting the statistical independence of observations from different sources and different instruments.

For the analysis of archival VHE spectra the observables are not raw counts but published, instrument-response-corrected flux points $d_{ik} \pm \sigma_{ik}$ taken from the STeVECat compilation. In this data mode the likelihood for dataset $k$ is Gaussian,
\begin{equation} \label{eq:gausslik}
\ln \mathcal{L}_k = -\frac{1}{2}\sum_i
  \frac{\left[d_{ik} - s_k\, m_{ik}(r, \Omega_\mathrm{M}, \delta\Gamma_{s(k)}, b_{s(k)})\right]^2}{\sigma_{ik}^2}
  - \sum_i \ln \sigma_{ik} + \mathrm{const}\, ,
\end{equation}
where $m_{ik}$ is the model flux of Eq.\,(\ref{eq:flux}) evaluated at the published energies, and $s_k$ is a per-dataset linear normalisation factor. Blazars are variable sources: the flux state during a given IACT campaign may differ by factors of a few from the seven-year \lat\ average that fixes the 3FHL anchor. If $s_k$ were held to unity this flux-state mismatch would be misinterpreted by the fit as a change in opacity, biasing the inferred opacity. We therefore allow one free normalisation per dataset, while keeping the spectral \emph{shape} anchored to the 3FHL parametrisation.%
We assign the proper bounded log-uniform prior
\begin{equation}
p(s_k)=\frac{1}{s_k\ln(s_{\max}/s_{\min})},\qquad
s_{\min}=10^{-4},\quad s_{\max}=10^{4},
\end{equation}
and marginalise it as
\begin{equation} \label{eq:freenorm}
\mathcal L_k^{\mathrm{marg}}=
\frac{1}{\ln(s_{\max}/s_{\min})}
\int_{s_{\min}}^{s_{\max}}\frac{\mathrm{d}s_k}{s_k}\,
\mathcal L_k(s_k)\,.
\end{equation}
The integral is evaluated with an interior Laplace approximation --- a Gaussian expansion about the interior maximum of the integrand, accurate here because the bounds lie more than two orders of magnitude beyond all fitted flux scales --- so that the dimensionality of the sampled parameter space remains that of the cosmological and hyperparameters alone, irrespective of the number of datasets. This robustness has a price: all cosmological information carried by the absolute flux level is discarded, and the constraint on $(r, \Omega_\mathrm{M})$ derives exclusively from the energy- and redshift-dependent \emph{shape} of the attenuation $e^{-\tau(E,z)}$ across each spectrum.
The fact that the spectral \emph{shape} of a source at the IACT epoch differs from its 3FHL anchor cannot be absorbed by $s_k$ and is modelled explicitly by the source-level hierarchy, whose recovery power is stress-tested in section~\ref{sec:intrinsicmismodel}.

Writing $u=\ln(E/E_{0,s})$, both anchor types are represented as
\begin{equation}
\frac{\mathrm{d}\Phi_{\mathrm{int},s}}{\mathrm{d}E}
\propto\exp(-\Gamma_su-b_su^2)\,.
\end{equation}
Thus $b_s=\beta_s$ for the natural-log 3FHL log-parabola and $b_s=0$ for a catalogue power law. The source index is shifted relative to its catalogue anchor according to
\begin{equation}
\Gamma_s=\Gamma_s^{\rm 3FHL}+\delta\Gamma_s,\qquad
\delta\Gamma_s=\mu_{\rm coh}+\sigma_{\Gamma,s}^{\rm 3FHL}\epsilon_s+\sigma_{\rm pop}t_s,
\quad \epsilon_s\sim\mathcal N(0,1),\quad t_s\sim t_{\nu=4}(0,1)\,.
\label{eq:indexhierarchy}
\end{equation}
The Student-$t$ population term permits a small number of GeV-to-TeV extrapolation outliers without allowing them to determine the common opacity; the choice $\nu=4$ gives tails heavy enough for such robustness while retaining a finite variance. This term with the standard deviation $\sigma_{\rm pop}$ enables us to account for the generally observed softening of the intrinsic spectral index from the GeV to the TeV band.
Curvature is treated as
\begin{equation}
b_s\sim\mathcal{TN}_{[0,\infty)}\!\left(
b_s^{\rm 3FHL}+\mu_b,
\sqrt{(\sigma_{b,s}^{\rm 3FHL})^2+\sigma_b^2}\right),
\label{eq:curvhierarchy}
\end{equation}
with $b_s^{\rm 3FHL}=0$ for catalogue power laws. Here $\sigma_{\Gamma,s}^{\rm 3FHL}$ and $\sigma_{b,s}^{\rm 3FHL}$ are the source-specific catalogue standard deviations. The notation $\mathcal{TN}_{[0,\infty)}(\mu,\sigma)$ denotes a normal distribution of untruncated mean $\mu$ and standard deviation $\sigma$, truncated at zero. The truncation encodes the standard assumption that intrinsic VHE spectra do not harden with increasing energy; an intrinsic upturn would moreover be nearly degenerate with a reduction of the opacity. A single $(\delta\Gamma_s,b_s)$ pair is shared by all datasets of source $s$. These source variables are integrated numerically by deterministic quadrature on fixed per-source grids in the index shift and the curvature, with the Student-$t$ population term represented as a finite Gaussian scale mixture, so only the cosmological and hyperparameters enter the sampler.

\subsubsection{Prior distributions and the opacity ratio}
The common source-population priors are
\begin{equation}
\mu_{\rm coh}\sim\mathcal N(0,0.11),\qquad
\mu_b\sim\mathcal N(0,0.04),\qquad
\sigma_{\rm pop}\sim\mathcal{HN}(0.30),\qquad
\sigma_b\sim\mathcal{HN}(0.14),
\label{eq:sourcepriors}
\end{equation}
where $\mathcal N(\mu,\sigma)$ denotes a normal distribution, $\mathcal{HN}(\sigma)$ a half-normal distribution of scale $\sigma$, and the catalogue uncertainties enter source by source in Eqs.~(\ref{eq:indexhierarchy}) and~(\ref{eq:curvhierarchy}). The prior scales were calibrated from deabsorbed 3FHL-to-VHE residuals in the $z<0.05$ subset, where the cosmological opacity is small; only the residual scales, not their measured signs, are used, so the coherent priors remain centred at zero. This empirical calibration uses sources that also occur in the full fit and should therefore be understood as an informative regularisation of the source hierarchy. The direct opacity parameters have broad priors
\begin{equation}
r\sim\mathcal U(0.2,3.0),\qquad
\Omega_\mathrm{M}\sim\mathcal U(0.05,0.70)\,.
\label{eq:cosmopriors}
\end{equation}

We adopt the baseline SL21 EBL photon density field without adjusting its underlying galaxy luminosity-density geometry across trial cosmologies. The effective optical depth scales directly via the dimensionless opacity parameter $\alpha_{\rm EBL}$ as $\tau(E,z) = \alpha_{\rm EBL} \mathcal{T}_{\rm SL21}(E,z;\Omega_\mathrm{M},h)$. Because the line-of-sight element $\mathrm{d}l/\mathrm{d}z$ carries a factor of $H_0^{-1}$, the overall normalisation scales with the identifiable ratio $r$, defined as:
\begin{equation}
r \equiv \frac{\alpha_{\rm EBL}}{h}\,.
\label{eq:rdef}
\end{equation}

To translate the direct posterior into a conditional $H_0$ posterior we use $h\sim\mathcal U(0.4,1.1)$ and an external truncated-normal EBL-scale distribution $p_\alpha$ on $0.3<\alpha_{\rm EBL}<2$. With $\mathcal{D}$ denoting the flux-point data and $p_\gamma$ the direct $\gamma$-ray posterior, the change of variables $\alpha_{\rm EBL}=rh$ gives
\begin{equation}
p(r,h,\Omega_\mathrm{M}\mid \mathcal{D},p_\alpha)
\propto p_\gamma(r,\Omega_\mathrm{M}\mid \mathcal{D})\,
h\,p_\alpha(rh)\,p_h(h),
\label{eq:conditionalh}
\end{equation}
where the factor $h$ is the Jacobian. We report the point-mass assumption $\alpha_{\rm EBL}=1$ and truncated-normal widths $\sigma_\alpha=0.10$ and 0.18. The latter is the median fractional half-width of the official SL21 optical-depth envelope~\cite{2021MNRAS.507.5144S} over $0.1<\tau<5$, $z\leq1$ and $0.1$--$20\,\mathrm{TeV}$, rounded from 0.19. A single scale treats this wavelength-dependent envelope as coherent; residual shape errors are tested separately in section~\ref{sec:eblmismodel}. These are alternative external assumptions applied to the same $p_\gamma(r,\Omega_\mathrm{M}\mid \mathcal{D})$, not additional fits of the $\gamma$-ray data.

\subsubsection{Posterior sampling}
The global posterior in $(r,\Omega_\mathrm{M},\mu_{\rm coh},\mu_b,\sigma_{\rm pop},\sigma_b)$ is explored with the reactive nested sampler UltraNest~\cite{Buchner:2021ultranest}, using 400 live points and population slice sampling. The production optical-depth response is precomputed on a 32-node $\Omega_\mathrm{M}$ grid and interpolated during inference; additional nodes are used only for validation. The production run required the posterior effective sample size (ESS) to exceed 400 and the estimated uncertainty in $\ln Z$ to be below 0.5; the realised values are ESS$=2683$ and $\sigma(\ln Z)=0.15$. Independent interpolation checks give a maximum relative error of $2.9\times10^{-4}$. Marginal results are quoted as the posterior median with the 16th and 84th percentiles. Two-dimensional $1\sigma$ and $2\sigma$ contours are kernel-density highest-posterior-density regions containing 68.3\% and 95.4\% of the posterior mass, respectively.
In what follows, they will be referred to as confidence intervals for simplicity.

\subsection{Systematic uncertainties}
Two classes of systematic uncertainties are considered beyond the statistical uncertainties entering the likelihoods of section~\ref{sec:likelihood}, in order to quantify the impact of the EBL modelling uncertainties and source spectrum modelling
on the determination of the cosmological parameters. Both are tested with matched injections: in every case the mock sample contains the same 241 datasets and 1922 energy points as the data sample and is generated at $(h,\Omega_\mathrm{M},\alpha_{\rm EBL})=(0.70,0.315,1)$. These are noise-free Asimov mocks: each flux is set to its model expectation while retaining the measured uncertainty of the corresponding archival point. Recovery uses the unchanged baseline hierarchy, so shifts relative to the matched control isolate the injected mismatch rather than a particular noise realisation.

\subsubsection{EBL mismodelling effect}
\label{sec:eblmismodel}
Given the presently limited accuracy of the EBL SED determination at $z=0$ from the various probes discussed in section~\ref{subsec:meas}~\cite{Hauser:2001xs,Keenan_2010,doi:10.1126/science.1258168} (see, for instance, figure 3 of Ref.~\cite{Baxter:2026tev}) we further explore energy-dependent systematic uncertainties in its 
spectral energy distribution impacting the derivation of the cosmological parameters.
To assess the impact of a residual EBL model error that is \emph{not} captured by the overall opacity scale $\alpha_{\rm EBL}$, we perform a dedicated mismodelling study. The mock observations are generated with a deliberately perturbed EBL model as the truth, and the recovery is performed with the unperturbed SL21 baseline. The perturbation tilts the spectral energy distribution in wavelength coordinate space as
  \begin{equation}
    \lambda I_\lambda^\mathrm{mod}(\lambda, z)
      = \lambda I_\lambda^\mathrm{ref}(\lambda, z)
      \left(\frac{\lambda}{\lambda_0}\right)^{\!\kappa},
    \label{eq:ebl_tilt_lambda}
  \end{equation}
where $\lambda_0 = 10\,\mu\mathrm{m}$ is the pivot wavelength, and positive (negative) values of $\kappa$ tilt the SED towards the mid- and far-infrared (the ultraviolet-to-near-infrared) side of the pivot. The values $\kappa = \pm 0.03$ represent a $\sim\!7\%$ change per wavelength decade, comparable to the SL21 shape uncertainty over the constraining range; the values $\kappa = \pm 0.25$ are retained as deliberately extreme stress tests (since $10^{0.25} \simeq 1.78$, they correspond to almost a factor-of-two change per decade). Such an approach enables us to capture the uncertainties in the 100 and 1\,$\mu$m peaks of the SED, respectively. Hereafter, this approach will be referred to as the \textit{tilt-shape} uncertainty.

As a second and complementary, more localised deformation we also consider a rescaling of the relative proportion of the two EBL peaks: the intensity of the cosmic infrared background (CIB) peak at $\lambda \sim 100$--$1000\,\mu\mathrm{m}$ is multiplied by a factor of $1.3$ through a smooth logistic transition centred on the COB/CIB valley at $\lambda \simeq 10\,\mu\mathrm{m}$, while the cosmic optical background peak at $\sim 1\,\mu\mathrm{m}$ is left at its baseline value. This mimics the case in which the dust-reprocessed component of the EBL, the part most affected by the zodiacal foreground subtraction, is systematically misestimated while the integrated-counts COB is correct. In both perturbation approaches the redshift evolution of the SED is left untouched, so that the deformation isolates the pure shape error of the local EBL template. No continuous tilt parameter is added to the production fit: an energy-dependent rescaling of the opacity is strongly degenerate with the cosmological energy dependence that carries the signal, and would act as an unconstrained absorber of the quantity being measured. Hereafter, this approach will be referred to as the \textit{ratio-shape} uncertainty.

\subsubsection{Intrinsic-spectrum mismodelling}
\label{sec:intrinsicmismodel}
The complementary modelling systematic concerns the intrinsic source spectra: any difference between the 3FHL anchor and the true intrinsic spectrum at the IACT epoch is removed neither by the per-dataset normalisation $s_k$ of Eq.~(\ref{eq:freenorm}), which absorbs only achromatic offsets, nor by the overall opacity scale, which acts coherently on all sources. The corresponding injections test precisely the degrees of freedom that the source hierarchy is intended to accommodate. We apply coherent index shifts over $-0.2\leq\Delta\Gamma\leq0.2$, representing a systematic calibration offset of the catalogue (including the absorption-induced softening discussed in section~\ref{sec:selection}), coherent curvature shifts $0.01\leq\Delta b\leq0.10$, representing spectral curvature at TeV energies that the GeV-band anchors cannot capture, and independent source-to-source index scatters of width $\sigma_\Gamma=0.10$ and $0.30$. The scatter cases use three antithetic realisation pairs at each width: within each pair the vector of source perturbations is sign-reversed, so that chance alignments of a particular random draw average out between the members of a pair. Each injected source perturbation is shared by all of its datasets, while the recovery is performed with the original 3FHL anchors and the priors of Eq.~(\ref{eq:sourcepriors}). 
These injections evaluate how effectively the hierarchical model recovers cosmological parameters under realistic spectral deviations, without adding ad hoc systematic error terms post-fit.

\section{Results}
\label{sec:results}
All results below use the full sample of 241 STeVECat datasets from 50 sources. The primary result is the direct posterior $p_\gamma(r,\Omega_\mathrm{M}\mid \mathcal{D})$; the values of $H_0$ quoted below are explicitly conditional transformations under the stated EBL-scale assumptions.

\subsection{Sensitivity and validation on mock observations}
The matched injection has $r_{\rm inj}=1/0.70=1.429$. It is recovered as $r=1.438^{+0.076}_{-0.075}$, while $\Omega_\mathrm{M}=0.372^{+0.217}_{-0.218}$ remains consistent with the nearly uniform input prior. With $\alpha_{\rm EBL}=1$, the corresponding conditional result is $H_0=70.0^{+4.1}_{-3.6}\,\mathrm{km\,s^{-1}\,Mpc^{-1}}$. For $\sigma_\alpha=0.18$ it becomes $72.3^{+13.2}_{-12.5}\,\mathrm{km\,s^{-1}\,Mpc^{-1}}$; the small upward shift of the median is induced by the transformation, its Jacobian and the finite $h$ prior rather than by the $\gamma$-ray likelihood.

\begin{figure}[!ht]
\centering
\includegraphics[width=0.72\linewidth]{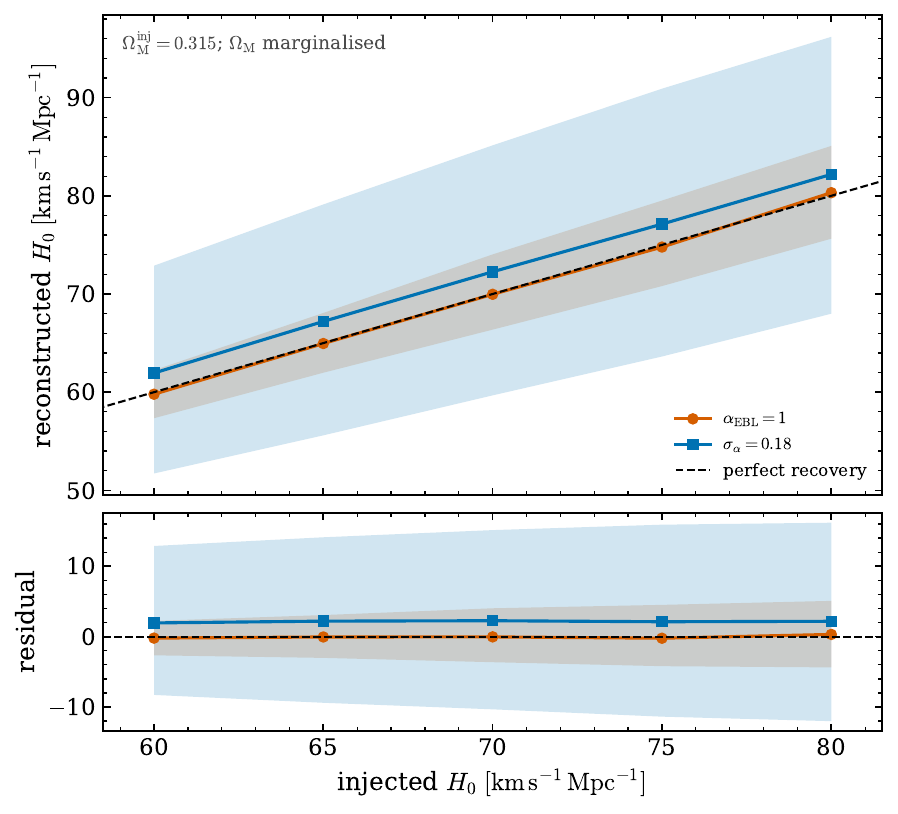}
\caption{Injection recovery for the noise-free Asimov sample of 241 datasets. The injected values are $H_0=60$, 65, 70, 75 and $80\,\mathrm{km\,s^{-1}\,Mpc^{-1}}$, with $\Omega_\mathrm{M}^{\rm inj}\equiv 0.315$ and $\alpha_{\rm EBL}^{\rm inj} \equiv 1$. In the recovery, $\Omega_\mathrm{M}$ is sampled over $0.05$--$0.70$ and marginalised. At each injected value, the points and bands show the median and central $1\sigma$ interval after transforming the corresponding direct $r$ posterior with $\alpha_{\rm EBL}=1$ (orange circles) or $\sigma_\alpha=0.18$ (blue squares). The dashed line denotes perfect recovery; the lower panel shows the residual.}
\label{fig:performanceplot}
\end{figure}

Figure~\ref{fig:performanceplot} shows the single injection-recovery scan used to validate the response over $H_0=60$--$80\,\mathrm{km\,s^{-1}\,Mpc^{-1}}$. For $\alpha_{\rm EBL}=1$, a linear fit has slope 1.017 and an rms departure of $0.20\,\mathrm{km\,s^{-1}\,Mpc^{-1}}$ from perfect recovery. The $\sigma_\alpha=0.18$ curve has slope 1.008 and the expected nearly constant conditional offset of about a few per cent. Thus the direct opacity inference is linear across the range relevant to the Hubble tension; widening the EBL-scale distribution broadens the result but does not change its response.

\subsection{Systematic-error budget from EBL and template mismodelling}
\begin{figure}[!ht]
\centering
\includegraphics[width=0.98\linewidth]{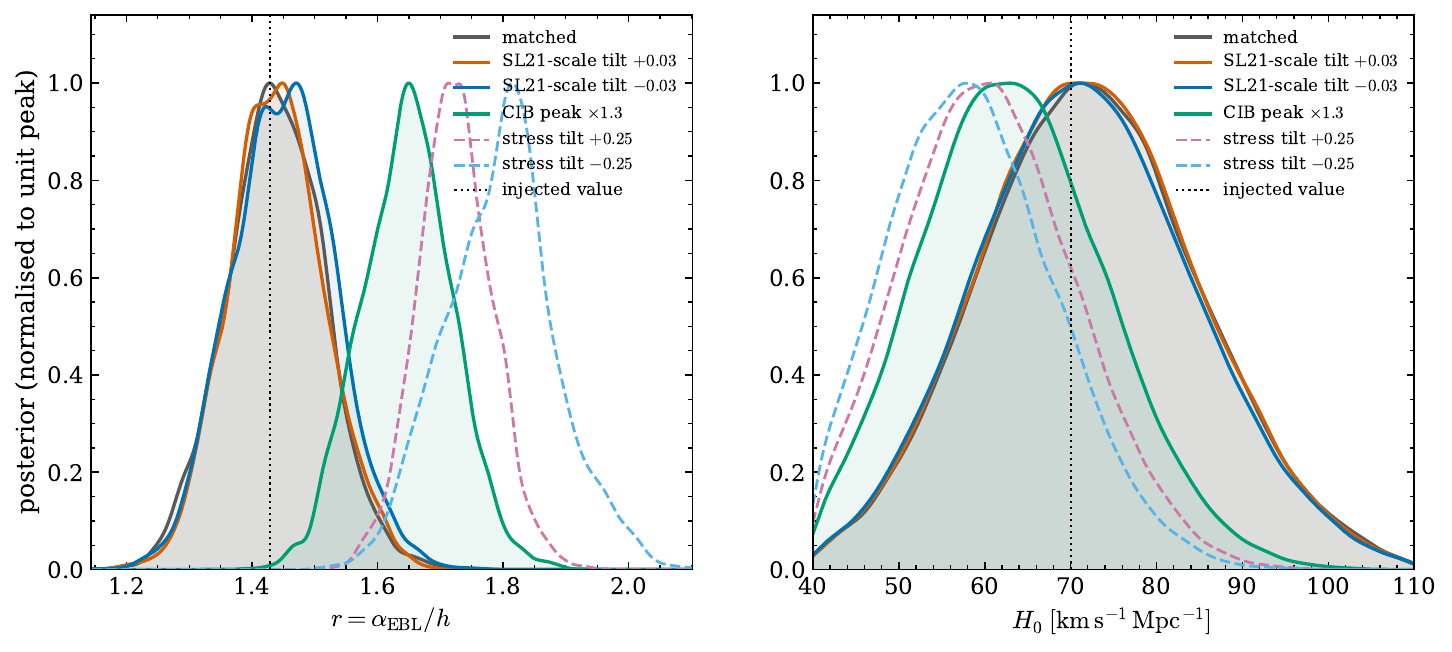}
\caption{EBL mismodelling recovery for the noise-free Asimov sample. All mocks have $(H_0,\Omega_\mathrm{M},\alpha_{\rm EBL})^{\rm inj}=(70\,\mathrm{km\,s^{-1}\,Mpc^{-1}},0.315,1)$; $\Omega_\mathrm{M}$ is sampled and marginalised in the recovery rather than fixed. \textit{Left panel:} direct posterior of $r=\alpha_{\rm EBL}/h$. \textit{Right panel:} conditional $H_0$ posterior for $\sigma_\alpha=0.18$. Grey denotes the matched control; solid curves show the SL21-scale tilts and the $1.3\times$ CIB peak, and dashed curves show the extreme $\kappa=\pm0.25$ stress tests. Dotted vertical lines mark the injected values.}
\label{fig:EBLmismodelling}
\end{figure}

The EBL recovery test is shown in figure~\ref{fig:EBLmismodelling}. The SL21-scale tilts $\kappa=\pm0.03$ shift the mean $r$ by only $+0.002$ and $+0.012$ ($+0.1\%$ and $+0.8\%$ relative to the matched control); for $\sigma_\alpha=0.18$, the corresponding conditional $H_0$ shifts are below $0.6\,\mathrm{km\,s^{-1}\,Mpc^{-1}}$ (0.8\%). Enhancing the CIB peak by 1.3 gives $\Delta r=+0.209$ ($+14.5\%$) and a conditional $H_0$ shift of $-8.9\,\mathrm{km\,s^{-1}\,Mpc^{-1}}$ ($-12.3\%$). The extreme tilts give $\Delta r=+0.285$ and $+0.357$ ($+19.8\%$ and $+24.8\%$ respectively). Their conditional $H_0$ shifts are $-11.7$ and $-13.7\,\mathrm{km\,s^{-1}\,Mpc^{-1}}$ ($-16.2\%$ and $-18.9\%$), respectively. These last cases are stress bounds and also produce artificial preferences toward the $\Omega_\mathrm{M}$ prior boundaries. Because the CIB rescaling contains a large coherent scale component, its full shift should not be added in quadrature to $\sigma_\alpha$; the figure is retained as a transparent sensitivity test of residual EBL-shape error. 

\begin{figure}[!ht]
\centering
\includegraphics[width=0.98\linewidth]{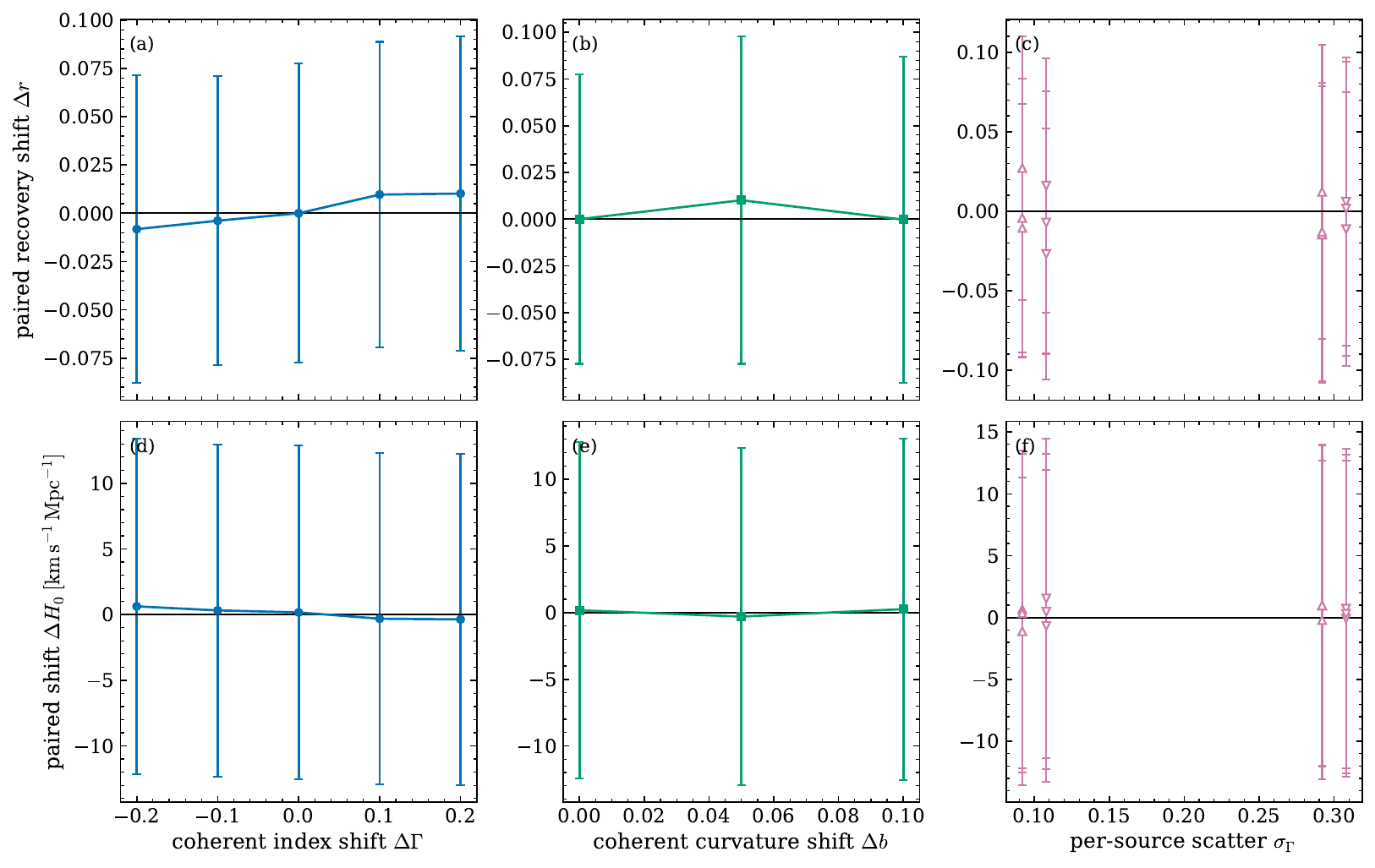}
\caption{Intrinsic-spectrum recovery power of the hierarchical model for the noise-free Asimov sample, injected with $(H_0,\Omega_\mathrm{M},\alpha_{\rm EBL})=(70\,\mathrm{km\,s^{-1}\,Mpc^{-1}},0.315,1)$. Recovery samples and marginalises $\Omega_\mathrm{M}$. The top row shows the shift of the posterior mean of the direct opacity ratio $r$ relative to the matched control; the bottom row translates the same runs into the conditional shift in $H_0$ for $\sigma_\alpha=0.18$. Columns show coherent index shifts (left), coherent curvature shifts (centre), and independent per-source index scatter (right). To give uniform sampling on the linear horizontal axes, the displayed coherent injections are $\Delta\Gamma=-0.20,-0.10,0,0.10,0.20$ and $\Delta b=0,0.05,0.10$; the denser completed scans are retained in the numerical analysis. Error bars give the posterior standard deviation in each mismodelled run; opposite triangles identify antithetic scatter realisations.}
\label{fig:intrinsicscatter}
\end{figure}

The source-population recovery in figure~\ref{fig:intrinsicscatter} illustrates the protection supplied by the hierarchy relative to a fixed-template analysis. Across the coherent index scan, the largest mean displacement is $|\Delta r|=0.010$ ($0.7\%$), corresponding to a conditional $H_0$ shift below $0.6\,\mathrm{km\,s^{-1}\,Mpc^{-1}}$ ($0.8\%$); coherent curvature shifts give at most $|\Delta r|=0.030$ ($2.1\%$), corresponding to approximately $1.5\,\mathrm{km\,s^{-1}\,Mpc^{-1}}$ ($2.1\%$) in conditional $H_0$. Both are small compared with the matched $\sigma(r)=0.077$ ($5.4\%$). Independent index scatter produces realisation-dependent shifts below 0.03 ($2.1\%$), with corresponding conditional $H_0$ shifts below $1.5\,\mathrm{km\,s^{-1}\,Mpc^{-1}}$ ($2.1\%$), and principally broadens the posterior: at $\sigma_\Gamma=0.30$ the typical $r$ width is about 18\% larger than in the matched case. The recovered population scales, rather than the cosmological ratio, accommodate the injected source mismatch. After transformation with $\sigma_\alpha=0.18$, the associated central-value shifts stay within approximately $1.5\,\mathrm{km\,s^{-1}\,Mpc^{-1}}$ ($2.1\%$), far below the conditional EBL-scale uncertainty.

\subsection{Application to the archival STeVECat source sample}
The hierarchical fit to the measured spectra yields
\begin{equation}
r=1.723^{+0.096}_{-0.095},\qquad
\Omega_\mathrm{M}=0.173^{+0.210}_{-0.092}.
\label{eq:realresult}
\end{equation}
The $1\sigma$ interval spans a substantial fraction of the $\Omega_\mathrm{M}$ prior, and the $2\sigma$ equal-tailed interval, $[0.054,0.600]$, is strongly prior-sensitive. We therefore do not interpret this posterior as a competitive matter-density measurement. Retaining the fiducial SL21 cosmology without recomputing its underlying luminosity-density geometry at each trial cosmology is sufficient for this predominantly low-redshift sample, as the induced geometrical shifts in the EBL photon field would be at the per-cent level across most of the grid and do not provide meaningful additional redshift leverage on $\Omega_\mathrm{M}$.

The source hierarchy is active in the real-data fit:
\begin{align}
\mu_{\rm coh}&=-0.064^{+0.070}_{-0.072}, &
\mu_b&=-0.026^{+0.034}_{-0.036},\nonumber\\
\sigma_{\rm pop}&=0.257^{+0.061}_{-0.049}, &
\sigma_b&=0.114^{+0.022}_{-0.019}.
\end{align}
The nonzero population widths show that the catalogue shapes should not be treated as exact; fixing them would fail to propagate this source-level uncertainty. Here the mismatch is included within the likelihood rather than appended as a separate error after the fit.

\begin{figure}[!ht]
\centering
\includegraphics[width=0.99\linewidth]{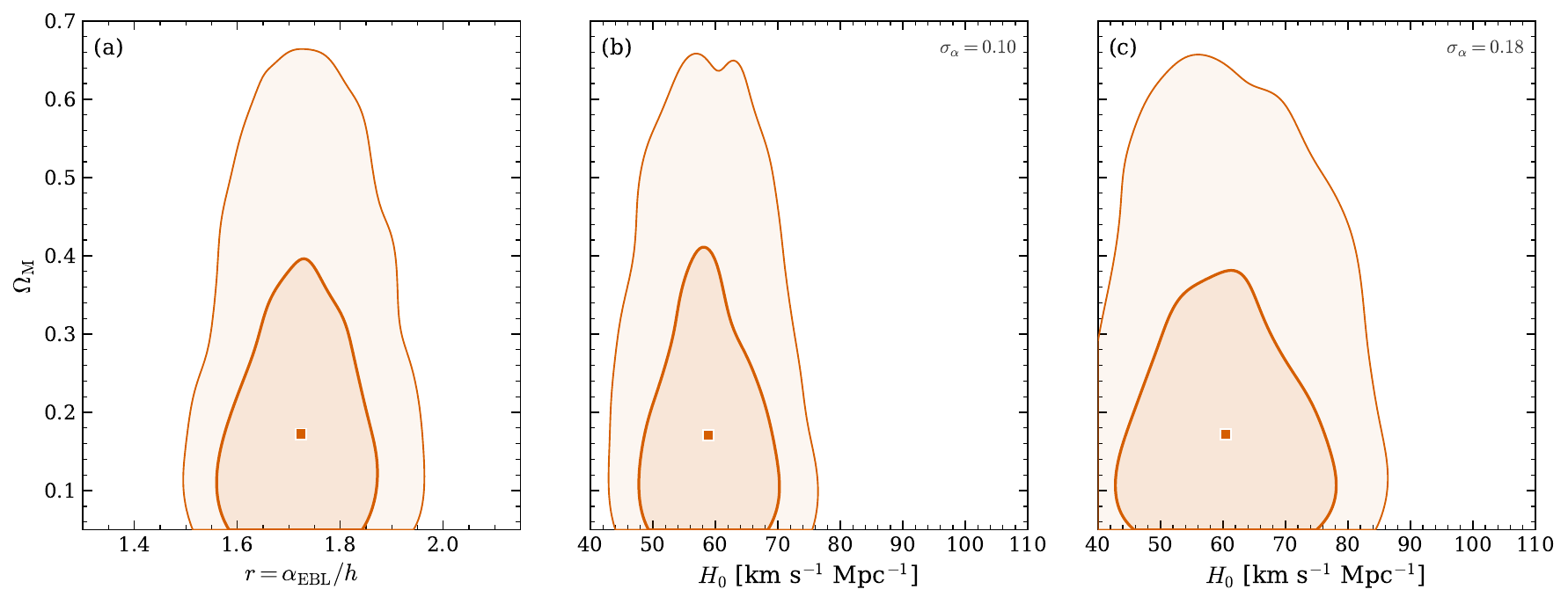}
\caption{Data-sample $1\sigma$ and $2\sigma$ contours. \textit{Left panel:} the direct $(r,\Omega_\mathrm{M})$ posterior. \textit{Middle and right panels:} conditional $(H_0,\Omega_\mathrm{M})$ posteriors obtained from the same direct posterior with $\sigma_\alpha=0.10$ and 0.18, respectively. Squares mark posterior medians.}
\label{fig:rhom_plane}
\end{figure}

The three declared EBL-scale scenarios give
\begin{align}
H_0&=58.3^{+3.4}_{-3.0}\,\mathrm{km\,s^{-1}\,Mpc^{-1}}, &&\alpha_{\rm EBL}=1,\nonumber\\
H_0&=58.8^{+6.8}_{-6.4}\,\mathrm{km\,s^{-1}\,Mpc^{-1}}, &&\sigma_\alpha=0.10,\nonumber\\
H_0&=60.5^{+10.8}_{-10.0}\,\mathrm{km\,s^{-1}\,Mpc^{-1}}, &&\sigma_\alpha=0.18.
\label{eq:h0scenarios}
\end{align}
Figure~\ref{fig:rhom_plane} keeps the likelihood result and its two finite-width transformations visually distinct. The left panel shows the measurement made by the $\gamma$ rays; the other panels show how external EBL-scale knowledge maps it into the $(H_0,\Omega_\mathrm{M})$ plane.

\subsection{Comparison with other determinations}
\begin{figure}[!ht]
\centering
\includegraphics[width=0.92\linewidth]{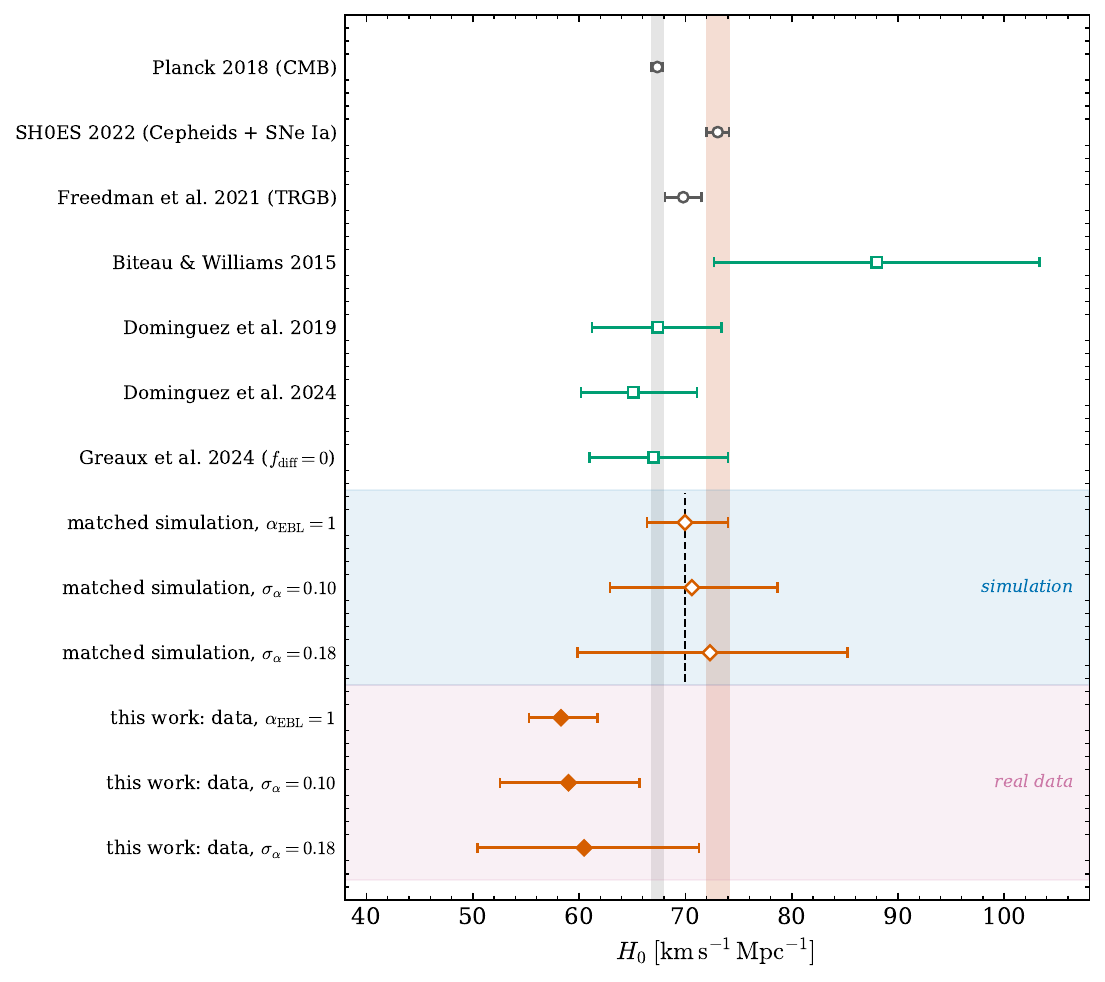}
\caption{Comparison of conditional $H_0$ results from the matched simulation and the data with representative early-Universe, late-Universe and previous $\gamma$-ray attenuation determinations~\cite{Planck:2018vyg,Riess:2021jrx,Freedman:2021ahq,Biteau:2015xpa,Dominguez:2019jqc,2024MNRAS.527.4632D,Greaux:2024coc}. The matched mock is injected with $(H_0,\Omega_\mathrm{M},\alpha_{\rm EBL})=(70\,\mathrm{km\,s^{-1}\,Mpc^{-1}},0.315,1)$. Open orange diamonds denote its three conditional recoveries ($\alpha_{\rm EBL}=1$, $\sigma_\alpha=0.10$, and $\sigma_\alpha=0.18$); the dashed vertical line spanning these points marks the injected $H_0$ value. Filled orange diamonds denote the $H_0$ determination from real data. Blue and purple shading distinguish the simulation and real-data results, respectively. Vertical bands show the Planck and SH0ES $1\sigma$ confidence bands.}
\label{fig:res_summary}
\end{figure}

Figure~\ref{fig:res_summary} compares the $H_0$ determination from the conditional one-dimensional results with representative determinations from Planck~\cite{Planck:2018vyg}, SH0ES~\cite{Riess:2021jrx}, the tip of the red-giant branch (TRGB)~\cite{Freedman:2021ahq}, and previous $\gamma$-ray attenuation based studies~\cite{Biteau:2015xpa,Dominguez:2019jqc,2024MNRAS.527.4632D,Greaux:2024coc}. The $\sigma_\alpha=0.18$ result is compatible with both sides of the Hubble tension but is much less precise; the fixed-scale value is lower. 

The lower fixed-scale value follows directly from the larger opacity ratio preferred by the archival spectra: for fixed $\alpha_{\rm EBL}$, $H_0=100\alpha_{\rm EBL}/r\,\mathrm{km\,s^{-1}\,Mpc^{-1}}$, and the measured $r=1.723$ therefore gives $H_0=58.3\,\mathrm{km\,s^{-1}\,Mpc^{-1}}$. The matched injections recover the input opacity without any significant bias. One possible contribution is the treatment of the intrinsic spectral shapes. In the real-data fit, the negative values of $\mu_{\rm coh}$ and $\mu_b$ favour spectra that are, on average, slightly harder and less curved than their 3FHL anchors, leaving more of the observed high-energy softening to be attributed to EBL attenuation. This comparison is not conclusive, however: Gr\'eaux et al.~\cite{Greaux:2024coc} marginalise independent curvature and cutoff parameters for each spectrum and include an energy-scale nuisance, whereas the Dom\'inguez et al. determinations~\cite{Dominguez:2019jqc,2024MNRAS.527.4632D} use previously reconstructed optical depths, including higher-redshift \lat\ measurements. While the data inputs and nuisance parameter treatments differ, the $H_0$ determinations are statistically compatible within a $1\sigma$ confidence interval.

\begin{figure}[!ht]
\centering
\includegraphics[width=0.74\linewidth]{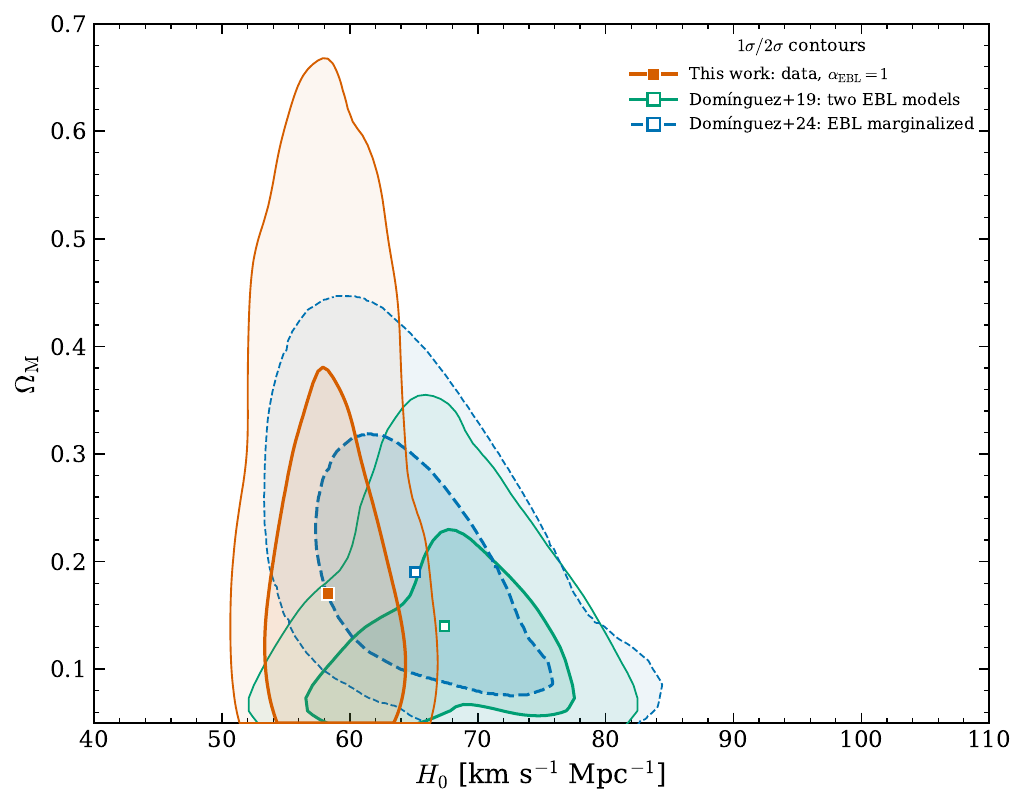}
\caption{Conditional data-sample $(H_0,\Omega_\mathrm{M})$ $1\sigma$ and $2\sigma$ contours for $\alpha_{\rm EBL}=1$, compared with the published Dom\'inguez et al. regions~\cite{Dominguez:2019jqc,2024MNRAS.527.4632D}, digitised from their figures. Dom\'inguez et al. (2019) is shown in green and Dom\'inguez et al. (2024) in blue; shaded regions mark their respective contour interiors. The overlay uses the closest available EBL-scale assumption, but the analyses differ in their opacity inputs and propagation of EBL and source-spectrum uncertainties.}
\label{fig:h0om_plane}
\end{figure}

For the $(H_0,\Omega_\mathrm{M})$ comparison, the fixed-scale case $\alpha_{\rm EBL}=1$ is the closest of our scenarios to analyses that do not fit a global EBL normalisation, and it is therefore the only one overlaid with the Dom\'inguez et al. contours in figure~\ref{fig:h0om_plane}. 
The results are compared while using different analysis methodologies.
Dom\'inguez et al.~\cite{Dominguez:2019jqc} propagated two EBL models and an opacity uncertainty, while Dom\'inguez et al.~\cite{2024MNRAS.527.4632D} marginalised over 500 SL21 EBL realisations. Both infer cosmology from previously measured optical depths, so their source-spectrum treatment is inherited from those measurements rather than fitted in the cosmological parameter determination.
Our hierarchy instead acts directly on the archival spectra, while holding the SL21 EBL field fixed to its fiducial model.
The tighter Dom\'inguez contours may come from larger statistical datasets with higher-z sample sources together with methodological and statistical framework choices.
The weak sensitivity to $\Omega_\mathrm{M}$ is consistent with the conclusions of Dom\'inguez et al.~\cite{Dominguez:2019jqc,2024MNRAS.527.4632D}, who find that the matter-density information is supplied primarily by the higher-redshift source observations. The present sample has a median source redshift of $z\simeq0.13$, where changes in $\Omega_\mathrm{M}$ have only a small effect on the line-of-sight opacity. Moreover, the many repeated spectra of nearby sources improve the precision on the common opacity ratio but do not extend the redshift lever arm. Once the leading opacity amplitude is described by $r$ and the per-dataset flux normalisations are marginalised, only the comparatively weak redshift-dependent change in the attenuation remains available to constrain $\Omega_\mathrm{M}$. A meaningful improvement therefore requires additional high-redshift sources and broader spectral coverage. Gréaux et al.~\cite{Greaux:2024coc} do not make use of a redshift-dependent model for the EBL SED but instead marginalise over the redshift dependence assuming multiple-Gaussian templates for the SED. From the 4FGL catalogue, the source spectral parameters are left free in their inference procedure with parameter priors incorporating reported catalogued values, while we instead allow for deviations in the spectral parameters in the population itself. While significant analysis choices differ, the $H_0$ determinations are compatible within $1\sigma$.

\section{Conclusions and outlook}
\label{sec:summary}

We have carried out an extensive analysis to measure the $\gamma$-ray opacity scale together with the cosmological parameters $(H_0, \Omega_\mathrm{M})$ from the EBL induced attenuation of selected VHE blazar spectra.
Our approach is deliberately conservative in its astrophysical assumptions: the intrinsic spectra are anchored to their \lat\ 3FHL parametrisations through a source-level hierarchical model with no injected cutoffs, blazar variability is absorbed by marginalised per-dataset normalisations, and the identifiability of the method is made explicit by the change of variables $r=\alpha_{\rm EBL}/h$: the $\gamma$-ray data mostly measure $r$, not $H_0$ and the EBL scale separately. Applied to 241 independent archival IACT datasets (energy spectra) from STeVECat, the method recovers injected opacity scales linearly and without appreciable bias, and yields $r=1.723^{+0.096}_{-0.095}$ from the real data. The conditional transformation gives $H_0=58.3^{+3.4}_{-3.0}\,\mathrm{km\,s^{-1}\,Mpc^{-1}}$ for a fixed SL21 EBL scale ($\alpha_{\rm EBL}=1$) and $H_0=60.5^{+10.8}_{-10.0}\,\mathrm{km\,s^{-1}\,Mpc^{-1}}$ for the SL21-motivated scale uncertainty $\sigma_\alpha=0.18$.

The performance studies on the reconstructed value of $H_0$
separate the error budget. Coherent index and curvature errors and population-level scatter in the intrinsic spectra are absorbed by the source hierarchy, with residual conditional $H_0$ shifts below about $1.5\,\mathrm{km\,s^{-1}\,Mpc^{-1}}$ in the tested cases. EBL-shape errors remain the more important external limitation: tilts of the size allowed by the SL21 uncertainty are subdominant, while a $1.3\times$ CIB enhancement shifts the conditional result by about $9\,\mathrm{km\,s^{-1}\,Mpc^{-1}}$ and deliberately extreme tilt stress tests give shifts of $12$--$14\,\mathrm{km\,s^{-1}\,Mpc^{-1}}$.
Keeping the SL21 baseline model fixed without dynamic cosmological rescaling is well justified for this predominantly low-redshift sample: the matter-density posterior remains broad and prior-sensitive, and a stronger constraint on $\Omega_\mathrm{M}$ requires additional redshift leverage and spectral coverage rather than a more restrictive intrinsic-spectrum model. The principal result of the present data analysis is therefore the direct opacity-scale ratio $r$ and its transparent transformation under alternative assumptions about the EBL scale.
The lower fixed-scale $H_0$ relative to previous $\gamma$-ray determinations is a consequence of the larger opacity ratio preferred in this analysis. The tendency of the hierarchy towards slightly harder and less curved intrinsic spectra provides one possible contribution, since it assigns more of the observed softening to EBL attenuation, but the difference cannot be isolated from the different data products and nuisance treatments used in the earlier analyses. The absence of a precise $\Omega_\mathrm{M}$ constraint has a more direct origin: the predominantly low-redshift source sample, including many repeated observations of the same nearby sources, provides precision on $r$ but little redshift leverage. 

No additional external prior is required for the direct determination of $r$. Its transformation into $H_0$, however, necessarily requires external information on the absolute EBL scale, which should remain explicit as in the scenarios considered here. A tighter prior on $\Omega_\mathrm{M}$ or more restrictive priors on the intrinsic source spectra would narrow the corresponding posteriors mainly by importing external information or limiting the allowed astrophysical freedom; they would not add redshift leverage to the present data. Improving the $H_0$ determination therefore requires a more precise external constraint on the EBL normalisation, whereas improving the $\Omega_\mathrm{M}$ determination requires observations at higher redshift rather than stronger priors.

The method devised to determine cosmological parameters using VHE $\gamma$-ray sources
is thus EBL-limited rather than data-limited: the $\gamma$-ray likelihood determines the opacity ratio to about $5.5\%$, so any external uncertainty on the absolute EBL scale larger than this dominates the conditional $H_0$ error budget. The forthcoming convergence of JWST galaxy counts with direct photometry, combined with the extended opacity lever arm of CTAO at higher redshifts and lower optical depths, will address this bottleneck directly. An external EBL-scale determination at the few-per-cent level would convert the sample analysed here into a $\sigma(H_0) \approx 3$--$4\,\mathrm{km\,s^{-1}\,Mpc^{-1}}$ measurement sufficient to address the Hubble tension with a probe fully independent of the distance ladder and the sound horizon determinations. A natural extension of the present framework, enabled by the analytic nuisance marginalisations that keep the sampled parameter space low-dimensional, is Bayesian model comparison between $\Lambda$CDM and evolving dark-energy cosmologies ($w$CDM, $w_0 w_a$CDM) via nested-sampling evidences, which is left for a forthcoming study. Another extension  would be to incorporate higher-redshift sources into the sample, possibly combining \lat and IACT measurements in the present Bayesian framework to assess the impact of systematic uncertainties in the ($H_0,\Omega_{\rm M}$) determination.

\acknowledgments
PL would like to thank Tomasz Łaguz for access to RTX 4090 and compute resources. This study was funded by ``Research Support Module'' as part of the ``Excellence Initiative -- Research University'' program at the Jagiellonian University in Kraków.

\bibliographystyle{JHEP}
\bibliography{bibliography,spectra_refs}

\end{document}